\documentclass[preprint,12pt]{elsarticle}
\usepackage{color}
\usepackage{graphicx,import}
\usepackage{subcaption}
\usepackage{epstopdf}
\usepackage{hyperref}
\usepackage{float}
\usepackage{mathrsfs}
\usepackage{multirow}
\usepackage{multicol}

\usepackage{amsmath,amsthm,amssymb}

\journal{Expert Systems With Applications}

\begin{document}

\begin{frontmatter}



\title{Aeroacoustic airfoil shape optimization enhanced by autoencoders}


\author[inst1]{Jiaqing Kou}
\ead{jiaqingkou@gmail.com}
\author[inst2]{Laura Botero-Bolívar}
\ead{l.boterobolivar@utwente.nl}
\author[inst1]{Román Ballano}
\ead{r.ballanom@alumnos.upm.es }
\author[inst1]{Oscar Marino}
\ead{oscar.marino@upm.es}
\author[inst2]{Leandro de Santana}
\ead{leandro.desantana@utwente.nl}
\author[inst1,inst3]{Eusebio Valero}
\ead{eusebio.valero@upm.es}
\author[inst1,inst3]{Esteban Ferrer}
\ead{esteban.ferrer@upm.es}

\address[inst1]{ETSIAE-UPM-School of Aeronautics, Universidad Politécnica de Madrid, Plaza Cardenal Cisneros 3, E-28040 Madrid, Spain}
\address[inst2]{Department of Thermal Fluid Engineering, University of Twente, PO Box 217, 7522 NB Enschede, The Netherlands}
\address[inst3]{Center for Computational Simulation, Universidad Polit\'ecnica de Madrid, Campus de Montegancedo, Boadilla del Monte, 28660, Madrid, Spain}



\begin{abstract}
We present a framework for airfoil shape optimization to reduce the trailing edge noise for the design of wind turbine blades. Far-field noise is evaluated using Amiet's theory coupled with the TNO-Blake model to calculate the wall pressure spectrum and fast turn-around XFOIL simulations to evaluate the boundary layer parameters. The computational framework is first validated using a NACA0012 airfoil at 0\textdegree $ $ angle of attack. Particle swarm optimization is used to find the optimized airfoil configuration. The multi-objective optimization minimizes the A-weighted overall sound pressure level at various angles of attack, while ensuring enough lift and minimum drag. 
We compare classic parametrization methods to define the airfoil geometry (i.e., CST) to a machine learning method (i.e., a variational autoencoder). We observe that variational autoencoders can represent a wide variety of geometries, with only four encoded variables, leading to efficient optimizations, which result in improved optimal shapes.
%
 When compared to the baseline geometry, a NACA0012, the autoencoder-based optimized airfoil reduces by $3\%$ (1.75 dBA) the overall sound pressure level (with decreased noise across the entire frequency range), while maintaining favorable aerodynamic properties in terms of lift and drag.
\end{abstract}




\end{frontmatter}


\tableofcontents

\section{Introduction}
\label{sec:intro}
Wind turbine noise has typically been one of the main drawbacks of the wide deployment and acceptance of wind devices for the generation of clean energy. The interest in controlling turbine acoustics is increasing since turbines are now being integrated in urban environments, where noise nuisances can be of great concern \cite{rogers2006wind,hubbard2009wind}. Wind turbine aeroacoustics is a complex phenomenon that includes several sources of noise. The most critical noise sources of a typical wind turbine are: 1) steady loading, which relates to the distribution of forces along the blade leading to broadband noise; 2) unsteady loading caused by the incoming sheared and turbulent atmospheric flow, associated with low frequency noise; and 3) airfoil self-noise, which encompasses various high frequency phenomena that relate to the boundary layer and eddies generated as the air passes the blades. Airfoil self-noise \cite{brooks1989airfoil} is the minimum amount of noise produced by an aerodynamic surface and is the main noise source in modern wind turbines~\cite{roger2005, oerlemans2011wind}. It is caused by the interaction of the turbulent boundary layer with the blade surface close to the trailing edge, when the hydrodynamic pressure fluctuations originated in the airfoil surface are scattered to the far field as noise, due to the sudden change of impedance in the blade trailing edge discontinuity~\cite{stalnov2016towards}. 
To generate quiet wind turbines, it is necessary to select airfoils that minimize noise generation in operational conditions (e.g. attached flow). Typical turbine airfoils were designed to be insensitive to roughness, by promoting the transition near the leading edge \cite{bertagnolio2001wind}, while noise generation was often neglected. However, social pressure to minimize wind farm noise leads to considering acoustics during the design phase of new airfoils.
The SIROCO project \cite{schepers2007sirocco,lutz2007design,Humpf_2007} incorporated aeroacoustic predictions for shape optimization design. Lutz et al. \cite{lutz2007design} proposed a method to predict airfoil trailing edge far-field noise, by combining XFOIL, a finite-difference code, and a modified TNO-TPD model. The method has been used to design new, less-noisy airfoils without loss in aerodynamic performance. Similarly, Hao et al. \cite{hao2008aerodynamic} introduced aerodynamic and aeroacoustic optimization of the wind turbine blade using a genetic algorithm. Lee et al. \cite{lee2013design} optimized the airfoils of the wind turbine based on genetic algorithms, to reduce airfoil self-noise, and validated the results by wind tunnel experiment. Rodrigues and Marta \cite{rodrigues2014addressing} performed multi-objective optimization to design wind turbine blades, where an increase in annual energy production of $15\%$ was achieved with a reduction in noise levels of $9.8\%$.  Zhou et al. \cite{zhou2015discrete} proposed a discrete adjoint framework for unsteady aerodynamic aeroacoustic optimization. More recently, Volkmer and Carolus
\cite{volkmer2018aeroacoustic} performed aeroacoustic airfoil shape optimization based on semi-empirical aeroacoustic models. There, XFOIL is combined with Amiet's model and evolutionary optimization to design airfoils with less noise while maintaining the required lift. Ricks et al. \cite{ricks2020cfd} proposed multi-objective aerodynamic-aeroacoustic shape optimization of airfoils based on a Reynolds-averaged Navier-Stokes solver with a state-of-the-art wall pressure spectrum model and Amiet’s model for trailing edge noise. Bu et al. \cite{yuepeng2020aerodynamic} developed the framework for aerodynamic/aeroacoustic variable-fidelity optimization of helicopter rotor based on hierarchical Kriging model. These works show the efficiency of using Amiet's theory and XFOIL for aeroacoustic optimization, whereas recent machine learning strategies (e.g., autoencoders) have not been considered.

Machine learning (ML) techniques are permeating in all fields of fluid dynamics \cite{yondo2018review,brunton_kutz_2019,GARNIER2021104973,kou2021data,vinuesa2022enhancing} and aerodynamic optimization; see the recent review by Li et al. \cite{li2022machine}.  In particular, unsupervised techniques for dimensionality reduction enable the treatment of problems with a large number of variables by coding the information into a reduced number of parameters (i.e., latent variables). Neural networks are ideal for representing nonlinear functions and can thus encode information very efficiently in very few latent variables.  Autoencoders (AEs) are artificial neural networks that can reduce the dataset information (encoding), but that can also reconstruct (decode) the original data, minimizing the reconstruction error. Variational autoencoders (VAEs) improve the behavior of classic autoencoders by enforcing latent variables to follow smooth statistical distributions (typically normal distributions) \cite{kingma_vae,zhao_vae}. By doing so, it is possible to avoid discontinuities in the latent variables, which helps to reconstruct the original database. Another advantage of requiring smooth latent variables is that they can be used for optimization. 

VAEs have not been widely adopted for optimization. In fact, to the authors' knowledge, only 
Rios et al. \cite{9504746Rios} have used VAEs for car optimization. They showed the superiority of this technique in incorporating local geometric features, when compared to modal techniques (e.g., principal component analysis). 
Additionally, Yonekura and Suzuki \cite{yonekura2021data} have shown the advantages of VAEs to represent airfoils using latent variables, but did not perform optimizations. 

Encouraged by the previous work, in this work we propose using VAEs to enhance the aeroacoustic optimization of wind turbine airfoils. We target the airfoil self-noise using Amiet's theory and develop an optimization framework that targets improved aerodynamics with minimum noise in the design.  We include VAEs for the parametrization of the airfoil geometry and compare this technique with the CST method \cite{kulfan2006fundamental} to parametrize the airfoil geometry. As shown in the results, VAEs show superiority and enhanced robustness over CST. 

The remaining of this paper is organized as follows. Section 2 introduces the far-field noise prediction framework, including the aeroacoustic and aerodynamic models, as well as the validation of the framework. The optimization framework is introduced in Section 3, where the shape parametrization including CST and VAEs, the optimization algorithm, and the problem formulation are detailed. The optimization results are summarized in Section 4. Conclusions are provided in Section 5.

\section{Far-field noise prediction}
\label{sec:method}
This section introduces the computational models used in the present study, including the aerodynamic and acoustic models. Amiet's theory~\cite{amiet1976TE} is adopted here for the trailing edge noise prediction. The theory calculates the far-field acoustic pressure spectrum using the wall pressure wavenumber-frequency spectral density in the vicinity of the trailing edge as input. This theory assumes a large span, a stationary observer and airfoil, and a uniform flow, and that the boundary layer turbulence is convecting over the trailing edge as a frozen pattern, i.e., the turbulence is not affected by the discontinuity of the trailing edge. 

\subsection{The acoustic model}
Equation~\ref{Eq:Amiet} presents the far-field power spectral density of an airfoil of chord \textit{c} and span \textit{b} for an observer perpendicular to the trailing edge at midspan (z\textsubscript{0} = 0)~\cite{stalnov2016towards}:

\begin{equation}\label{Eq:Amiet}
    S_{pp} (x_o, y_o, z_o = 0, \omega) = \left(\frac{\omega c y_o}{4\pi c_o \sigma^2} \right)\frac{b}{2} \left|\mathscr{L}(\kappa_x,\kappa_z = 0, x,y,U_\infty,\overline{U_c})\right|^2 \Lambda_{z|PP} \Pi_{\mathrm{pp}}(\omega).
\end{equation}

In this equation, $\omega$ is the angular frequency ($= 2\pi f$), and $x_o, y_o, z_o$ is the observer location. The reference system adopts $x$ in the streamwise direction, $y$ in the direction perpendicular to the wall and $z$ in the spanwise direction, with $x=y=z=0$ at the trailing edge of the airfoil in the mid-span. $c_o$ is the speed of sound, $\kappa_x$ and $\kappa_z$ are the spatial wavenumber in the $x$ and $z$ directions, and $\overline{U_c}=0.6\times U_\infty$ is the mean convection velocity. $\mathscr{L}$ is the airfoil response function proposed by Amiet~\cite{amiet1976TE} which is the transfer between hydrodynamic wall pressure fluctuations caused by the turbulent boundary layer to acoustic waves in the far field. Amiet considered a flat plate, therefore, $\mathscr{L}$ is constant among the different airfoils. The implementation of $\mathscr{L}$ followed in this work, was proposed by Roger et al.~\cite{roger2005} and considers the backscattering effect of the leading edge. $\Lambda_{z|PP}$ is the spanwise correlation length, which is related to the size of the vortices in the spanwise direction. Finally, $\Pi_{\mathrm{pp}}(\omega) $ is the wall pressure spectrum in the vicinity of the trailing edge and is comparable to the measurements of a sensor located on the airfoil surface. The wall-pressure spectrum is the only quantity that is different for each airfoil facing the same inflow conditions. The wall-pressure spectrum is calculated using the TNO-Blake model, explained in detail in~\ref{sec: WPS}. Additional details on how to compute the various parameters in Equation~\ref{Eq:Amiet}, are detailed in \ref{sec:appendix_amiet_param}.

\subsection{Aerodynamic simulation based on XFOIL }
Numerical simulations using XFOIL~\cite{drela1989xfoil} are conducted to obtain the boundary layer characteristics, see \ref{sec:appendix_BL_parameters}, for each candidate airfoil during the optimization procedure. XFOIL is a viscous-inviscid iterative software code that does not require any prior mesh preparation. The inviscid pressure distribution is modeled using a linear vortex strength distribution. Viscous effects and the development of the laminar-turbulent boundary layer are modeled using linear stability (i.e., the e method) and integral boundary layer theory. 
The XFOIL outputs are used to calculate the wall pressure spectrum, which, in turn, is used to predict the far-field noise produced by the airfoil. 

In all the cases included in this work, we fix the boundary layer transition near the leading edge in both pressure and suction sides, to mimic the insensitivity to roughness that is commonly sought in wind turbine airfoils. The number of panel nodes is set to 160 and the number of iterations to 1000. 

\subsection{Methodology validation: NACA0012 airfoil}
\label{sec:validation}
To validate the trailing edge noise prediction methodology, a widely studied airfoil in the literature, i.e., a NACA~0012 airfoil is used, comparing our results with those obtained in the literature. The airfoil is assumed to have a $0.4 m$ chord and a $1m$ span. The far-field observer is located at $1 m$ from the airfoil trailing edge at mid-span, i.e., $x = 0$, $y = 1m$, and $z = 0$. The wall pressure spectrum is calculated at $x/c~=~0.99$. The boundary layer transition is fixed at $x/c~=~0.06$ on both sides. The angle of attack is fixed at 0\textdegree. The inflow velocity is assumed to be $56~m/s$, which corresponds to a Reynolds number of 1.5~$\times$~10\textsuperscript{6} and a Mach number of $0.16$. The frequency resolution is 1~Hz and the far-field noise $S_\mathrm{pp}$ is calculated from 100~Hz to 10~kHz. Sound pressure level (SPL) is calculated at the observer location by

\begin{equation}\label{Eq:SPL}
    SPL = 20 log_{10}\frac{p}{p_{ref}},
\end{equation}

\noindent where ${p_{ref}}$ is the acoustic pressure reference, that for air, is $20~\mu Pa$, and $p$ is the far field pressure, obtained from the sum of the power spectral density ($S_{pp}$ from Eq.~\ref{Eq:Amiet}) from both, pressure and suction side, and multiplied by the frequency resolution.

\begin{figure}[htbp]
    \centering
    \includegraphics[width=0.9\linewidth]{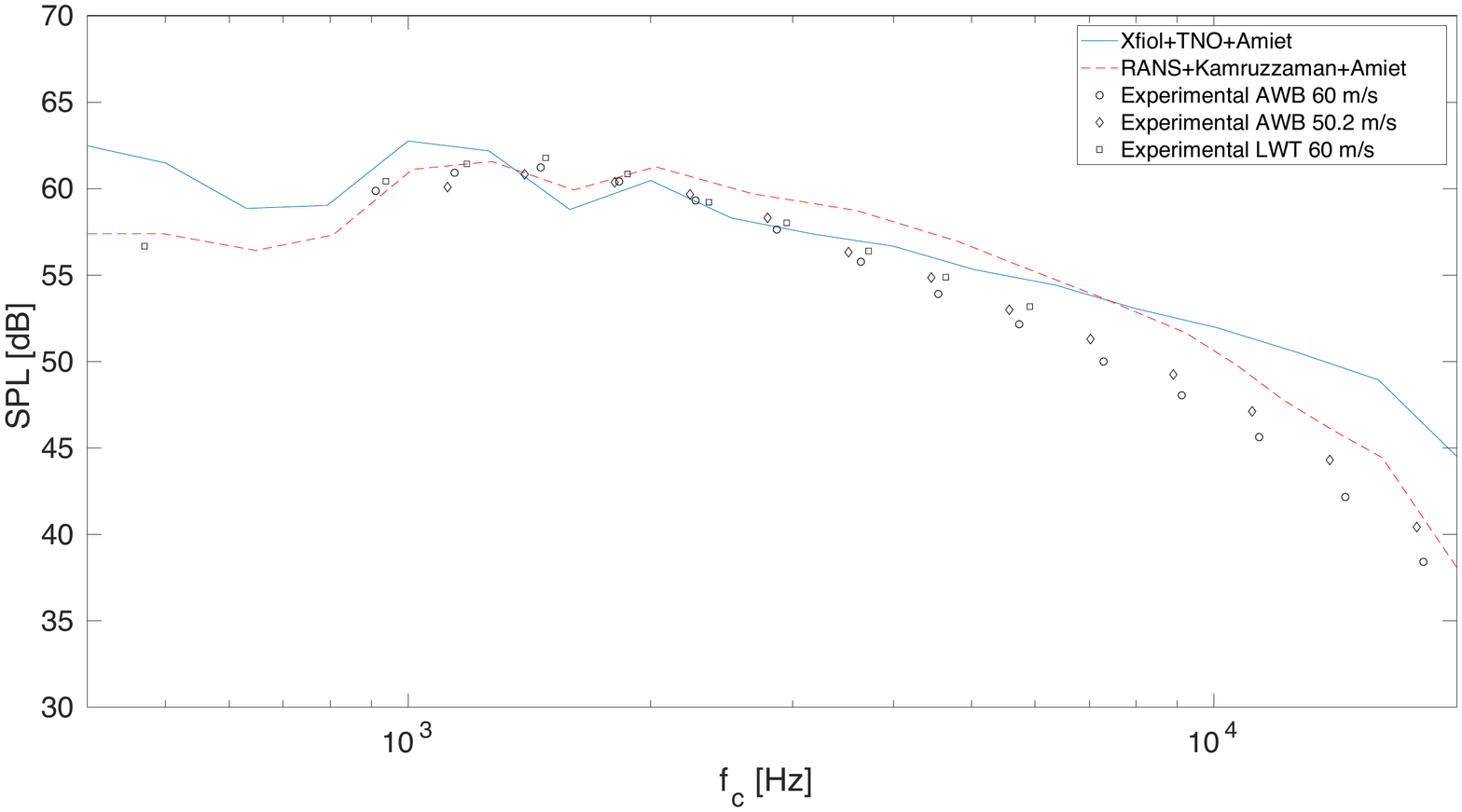}
    \caption{Far field Noise comparison between current methodology, RANS simulations (from~\cite{kuccukosman2018}) and experimental measurements (from~\cite{dlr78655}) for a NACA0012. Sound pressure levels are computed from Eq. \ref{Eq:SPL} and presented in $1/3$ octave band.}
    \label{fig:validation}
\end{figure}

Comparisons have been conducted with experimental results reported in~\cite{dlr78655} (results have been scaled to match the presented conditions) and with noise prediction using Amiet's theory coupled with the semi-empirical model proposed by Kamruzzaman et al.~\cite{kamruzzaman2015semi} and RANS simulations~\cite{kuccukosman2018}. Figure~\ref{fig:validation} shows the far-field noise spectrum in 1/3~octave for the predicted far-field using the methodology presented in this paper and in the literature. Additionally, Table~\ref{table:OasplValidation} shows the integrated pressure level of the spectrum for the same cases, between a frequency range of 1000 to 6300 Hz, i.e., the frequency range where all data is available. The results suggest that the applied methodology, i.e., XFOIL coupled with the TNO-Blake model and Amiet's theory, is appropriate to predict trailing edge noise and agrees with the literature showing less than 1~dB difference in the integrated spectra. Validation indicates that the proposed method shows comparable accuracy with more costly flow solvers, while still fast enough to be used for optimization where multiple evaluations of acoustic performance are needed for different airfoils.

\begin{table}[h!]
\caption{Comparison of the overall sound pressure level in the frequency range from 1000 to 6300 Hz. Values are presented in dB. (LWT: Laminar Wind Tunnel; AWB: Braunschweig Aeroacoustic Wind Tunnel)}
\label{tab:BC}
\label{table:OasplValidation}
\centering
\begin{tabular} {p{2cm}p{2cm}p{2cm}p{2cm}p{2cm}} 
 \hline
{Current} & {RANS \cite{kuccukosman2018}} & {AWB~\cite{dlr78655}} & {AWB~\cite{dlr78655}} & {LWT~\cite{dlr78655}}\\
{work} & {} & {60 m/s} & {50.2 m/s} & {60 m/s}\\
 68.87 & 68.81 & 68.36 & 67.67 & 68.83 \\ \hline
\centering
\end{tabular}
\end{table}

\section{Optimization framework}
\label{sec:ptimization}
The optimization framework is introduced, including the parameterization of the shape, the optimization algorithm, and the formulation of the problem (objective function and constraints). As a key ingredient to enhance aeroacoustic shape optimization, a variational autoencoder is introduced to parameterize the airfoil surfaces.

\subsection{Airfoil shape parametrization}
The review by Li et al. \cite{li2022machine} provides an overview of the techniques for airfoil shape parametrization. Our aim in this work is to assess the suitability of variational autoencoders for shape parametrization when performing optimization. To show the potential of autoencoders, we compare optimization using classic well-established CST parametrization with a variational autoencoder parametrization technique.   
\subsubsection{CST parametrization}
The Class/Shape function Transformation (CST), proposed by Kulfan and Bussoletti \cite{kulfan2006fundamental} in 2006, is a classic and commonly used method to represent/parameterize airfoils. CST describes the two-dimensional airfoil (with chord length $c$) as the product of a class function $C(x/c)$ and a shape function $S(x/c)$, plus a term that considers the trailing edge thickness ($\Delta z_{te}/c$) \cite{ceze2009study,sripawadkul2010comparison}:

\begin{equation}
    \frac{y}{c} = C\left(\frac{x}{c}\right)S\left(\frac{x}{c}\right) + \frac{x}{c}\frac{\Delta z_{te}}{c},
\end{equation}
where the class function $C(x/c)$ is given by:

\begin{equation}
    C\left(\frac{x}{c}\right) = \left(\frac{x}{c}\right)^{N_1}\left(1 - \frac{x}{c}\right)^{N_2}, \ 0 \leq \frac{x}{c} \leq 1.
\end{equation}

The exponents $N_1$ and $N_2$ are set to $0.5$ and $1.0$ for a general airfoil shape. This ensures a round leading edge ($\sqrt{x/c}$) and a sharp trailing edge $(1-x)$. To generate arbitrary airfoils, a Bernstein polynomial is used for the shape function:

\begin{equation}
    S\left(\frac{x}{c}\right) = \sum_{i=0}^n \left [ b_i \cdot K_{i,n} \cdot \left(\frac{x}{c}\right)^{i} \cdot \left(1 - \frac{x}{c} \right)^{n-i} \right],\ K_{i,n} = \frac{n!}{i!(n-i)!},
\end{equation}
where $b_i$ is the design parameter of the CST method. To generate smooth and well-defined airfoil configurations, the $x$ coordinates are spaced using conventional cosine clustering, with $200$ coordinates in total.

\subsubsection{Autoencoder-based parametrization}
\label{sec:autoencoder}
As a nonlinear dimensionality reduction technique, autoencoders have been used for many engineering applications, such as uncertainty quantification \cite{YONG2022118196}, object detection \cite{SCARPINITI2022116366}, turbulent flow prediction \cite{eivazi2022towards}, etc. Variational autoencoders have been introduced as a new parametrization method, which can enhance the parametrization of airfoils \cite{yonekura2021data}. VAEs allow for efficient random and guided sampling of novel airfoils from the lower-dimensional latent space of variables, avoiding nonphysical shapes, while allowing for a wide variety of shapes. 

\begin{figure}[htbp]
    \centering
    \includegraphics[scale = 0.7]{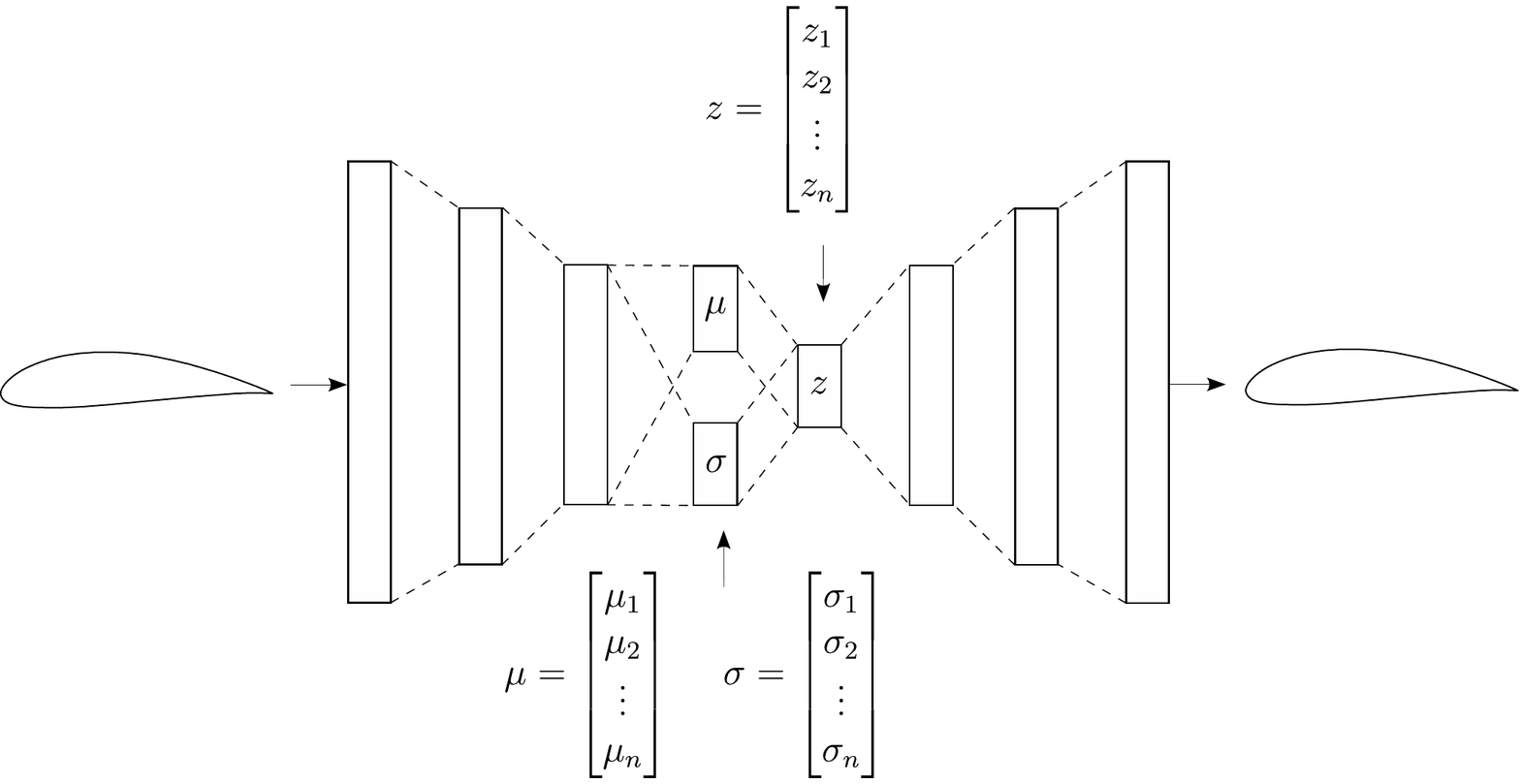}
    \caption{Architecture of the variational autoencoder.}
    \label{fig:archVAE}
\end{figure}

The proposed VAE has been trained using the UIUC airfoil database~\cite{UIUC_DB}, using 198 cosine-spaced points per airfoil. Since the $x$ spacing for every airfoil is the same, only the vertical coordinate $y$ is needed to train the network. The architecture (figure \ref{fig:archVAE}) consists of a symmetrically fully connected multilayer perceptron architecture (MLP) with a variable number of latent features $n$ between the encoder and decoder parts. The resulting network is capable of generating airfoil shapes by decoding an input set of $n$ latent variables, as shown in figure \ref{fig:archVAE}. By condensing the information from the original airfoil database into a lower-dimensional latent space, this allows for faster optimization iterations by means of dimensionality reduction. 

\begin{figure}[htbp]
    \centering
    \includegraphics[scale = 0.6]{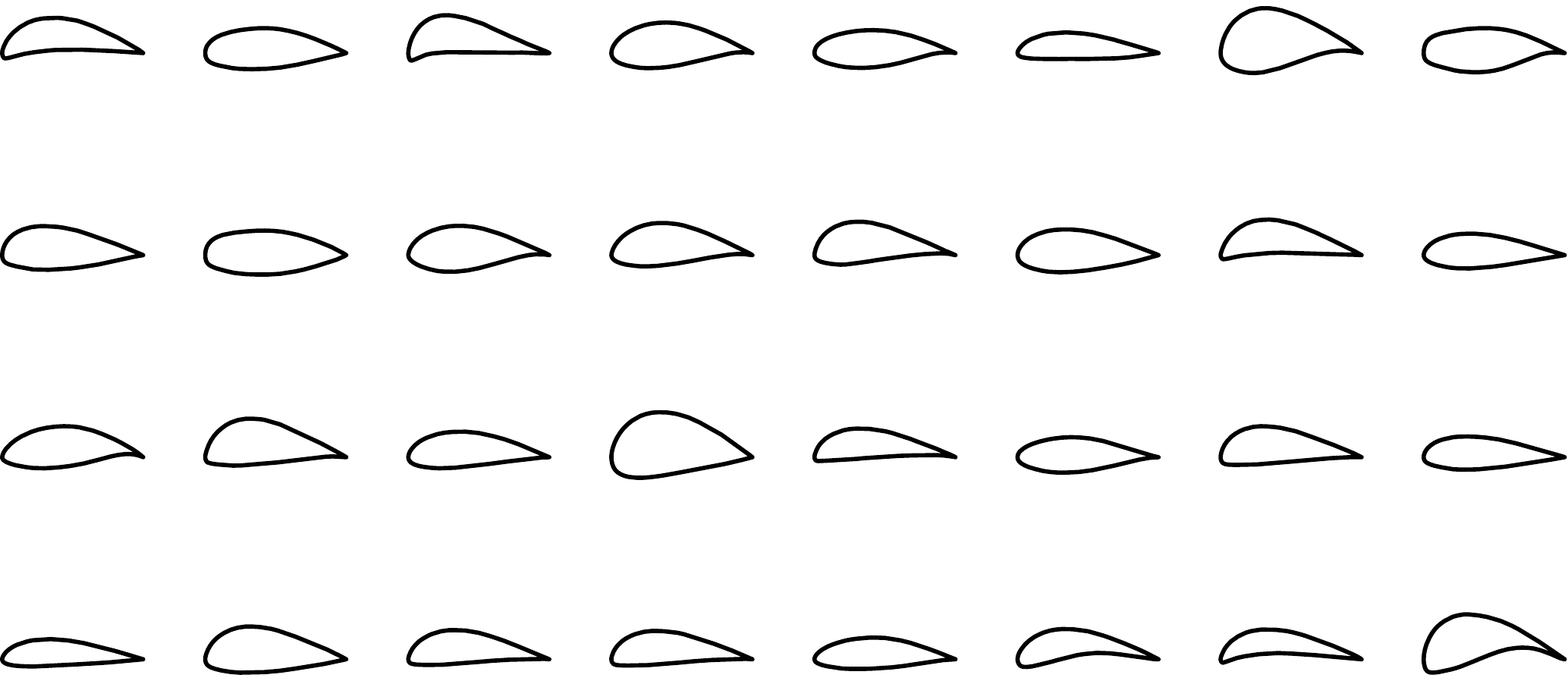}
    \caption{Randomly generated airfoils using a variational autoencoder ($\sigma = 1$, $n=4$).}
    \label{fig:randomsample}
\end{figure}


Latent variables contain valuable geometric information about the airfoil coordinates. For example, using only $n=4$ latent variable generator, we can obtain a large family of airfoils, as shown in Figure \ref{fig:randomsample}. The loss function $\mathcal{F}$, which is the sum of the reconstruction loss and the Kullback-Leibler (KL) divergence; see Equation \ref{eq:lossfunct} can be minimized through training. The reconstruction loss quantifies the mean square error (MSE) between the input and output coordinates, while the KL divergence tries to approximate the latent space to a standard normal distribution~\cite{kingma_vae}. In addition, a scaling term $k$ has been included in the KL divergence part of the loss function to reduce the influence of this term. Experimenting, we find an optimal value of $k=5\cdot 10^{-6}$, see \ref{sec:appendix_AE_parameters}.

\begin{equation}
\begin{gathered}
    \mathcal{F} = \mathcal{F}_{rec} + k\cdot D_{KL}(p\hspace{1mm}||\hspace{1mm}q) \\
    \mathcal{F}_{rec} = \left \| \mathbf{y_{out}}^2 - \mathbf{y_{in}}^2\right \| \\
    D_{KL}(p\hspace{1mm}||\hspace{1mm}q) = - \frac{1}{2}\sum_{i=1}^{n}\left ( 1+ log(\sigma_i^2) - \mu_i^2 - \sigma_i^2\right ).
 \end{gathered} \label{eq:lossfunct}
\end{equation}


A normally distributed latent space allows for random sampling of airfoils from a standard normal distribution $\mathcal{N}(0,1)$ in an efficient way, generating realistic airfoils. Control over the bounds of the latent variables is done by modifying the standard deviation $\sigma$ of the normal distribution. Namely, $\sigma>1$ generates airfoil shapes that are more likely to be outside the training envelope, while $\sigma<1$ tends to produce more conservative airfoil shapes. A parametric study is included in \ref{sec:appendix_AE_parameters} to help to determine the optimal autoencoder model for aeroacoustic optimization.

\subsection{Variational autoencoder parameters}\label{sec:appendix_AE_parameters}
In the following sections, we tune the variational autoencoder parameters to make it suitable for optimization while enhancing its performance. In particular, we study the number of latent variables and the value of the loss scaling parameter.

\subsubsection{Evaluation metric}
The metric used to evaluate the performance of the model is the Fréchet distance (FD), frequently used to assess image generation models, such as Generative Adversial Networks (GANs)~\cite{heusel_fid}. In the variational autoencoder, it is used to quantify the distance between the generated airfoil distribution, with normal distribution ($\mathbf{\mu_1}, \mathbf{\sigma_1}$) and the original airfoil distribution from the UIUC database, with normal distribution ($\mathbf{\mu_2}, \mathbf{\sigma_2}$):

\begin{equation}\label{Eq:frechet_distance}
 FD = d^2 = \left|\left|\mathbf{\mu_1} - \mathbf{\mu_2} \right|\right|^2 + Tr\left(\mathbf{\sigma_1} + \mathbf{\sigma_2} - 2\sqrt{\mathbf{\sigma_1}\cdot \mathbf{\sigma_2}}\right)
\end{equation}
where $\mathbf{\mu_i}$ is the mean vector and $\mathbf{\sigma_i}$ is the covariance matrix for each distribution. $Tr$ refers to the trace operation. A low value of FD indicates closeness between the source and generated airfoil distributions and, therefore, a good generative capacity.

\subsubsection{Loss scaling parameter k}
\begin{figure}[htbp]
    \centering
    \begin{subfigure}[]{\textwidth}
     \centering
     \includegraphics[width=0.8\textwidth]{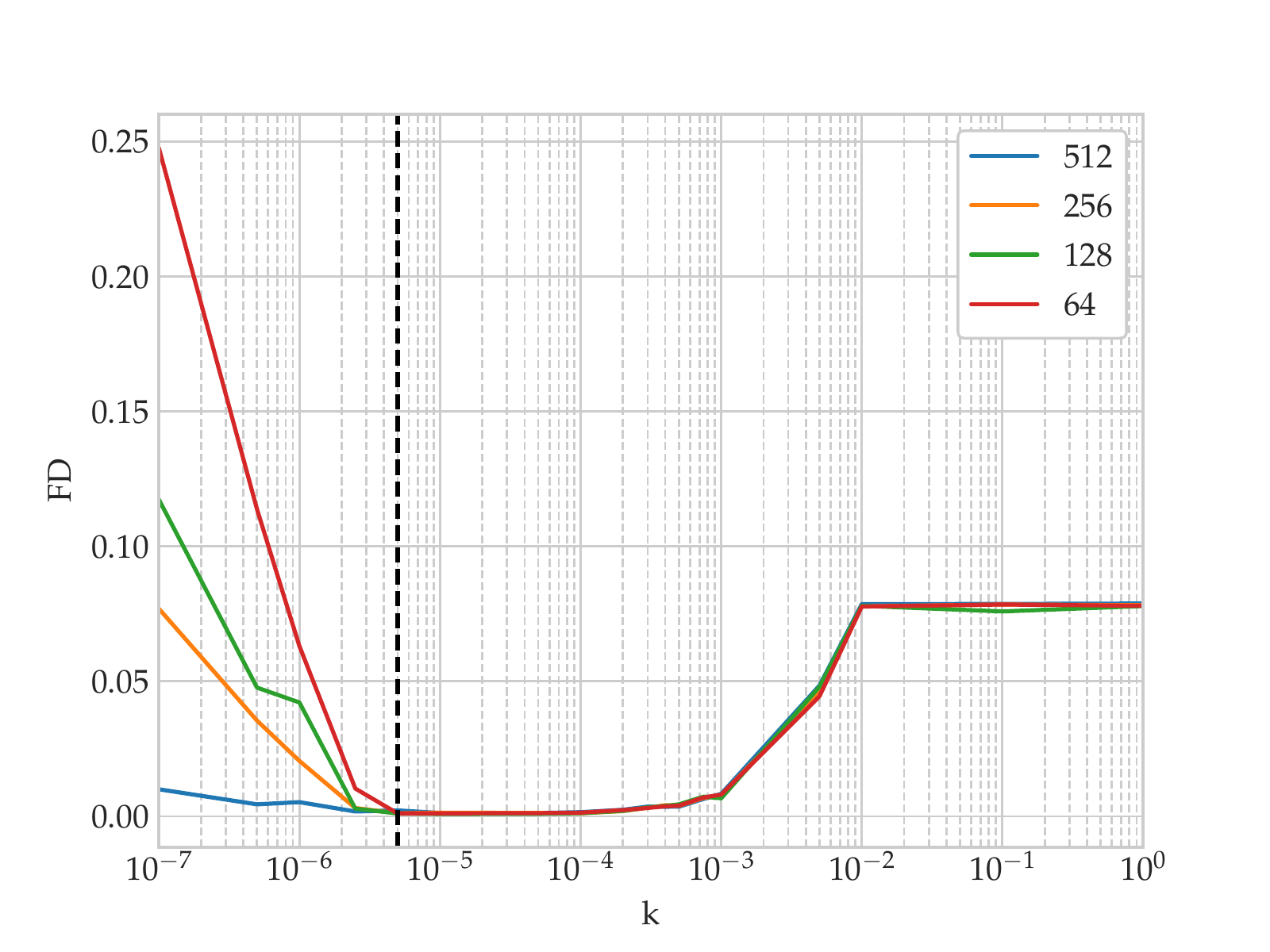}
     \label{fig:FIDvsk}
    \end{subfigure}

    \begin{subfigure}[]{\textwidth}
     \centering
     \includegraphics[width=0.8\textwidth]{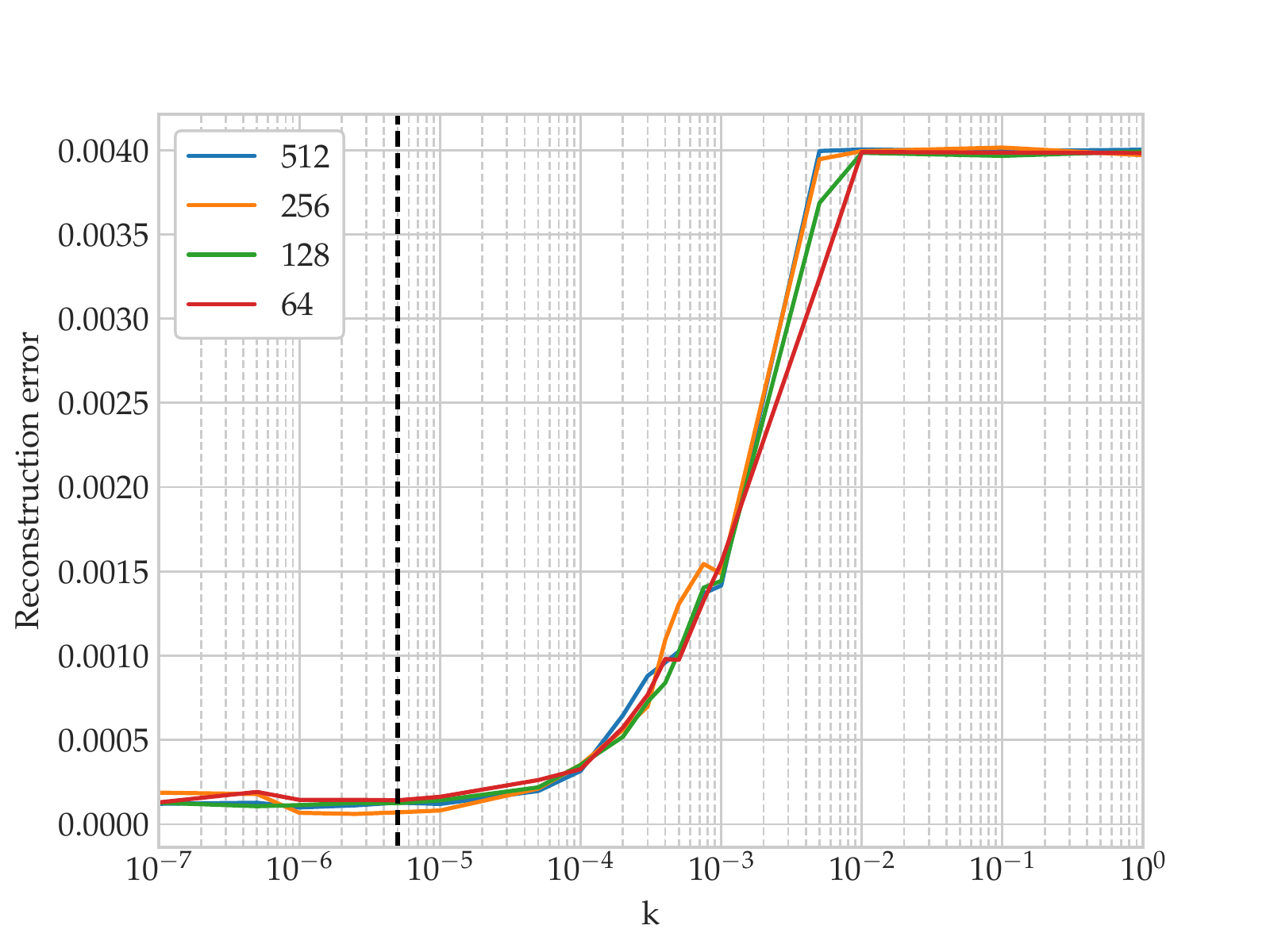}
     \label{fig:reconvsk}
    \end{subfigure}

    \caption{Fréchet distance (top) and reconstruction error (bottom) against $k$. Legend indicates number of hidden neurons in the first layer. Optimal $k = 5\cdot10^{-6}$ marked with dashed line.} \label{fig:kvariation}
\end{figure}

Equation \ref{eq:lossfunct} includes a loss scaling parameter k, that requires tuning. We study several base architectures consisting of three hidden layers where the number of hidden neurons is successively halved (512, 256, 128, 64) and depict the results in Figure \ref{fig:kvariation}. We set $n=4$ latent variables and use an encoder and decoder with mirrored structure. 

Figure \ref{fig:kvariation} shows a limited effect when changing the number of hidden neurons, both for the Fréchet distance (top) and the reconstruction error (bottom). The figure also shows the importance of selecting the loss scaling parameter $k$. 
For $k>10^{-2}$, the Kullback-Leibler divergence term dominates the loss function and causes the network to always output the same airfoil for any given input. This results in a stagnation in both the FD and the reconstruction error for large $k$. For low values of the scaling term, approximately $k<4\cdot10^{-6}$, the Fréchet distance (top figure) shows a significant decrease in generative quality. Indeed, when k is very low, the variational autoencoder resembles a regular autoencoder, in which latent features do not form a continuous space to sample from (i.e., not suitable for optimization). In this case, the randomly sampled airfoils may show nonphysical shapes, deviating from the statistics of the original dataset. For this reason, we select $k = 5\cdot 10^{-6}$ after evaluating the Fréchet distance, along with the reconstruction error.

\subsubsection{Autoencoder architecture}
We fix the variational autoencoder parameters to $k = 5\cdot 10^{-6}$ and optimize the number of neurons and layers for the latent variables $n=4$. To do so, we optimize the number of neurons and layers using Bayesian methods implemented in the Python package \textit{Optuna}. The algorithm employs a tree parzen estimator sampler along with a hyperband pruner. After 500 optimization iterations, the best architecture found is a mirrored three-layer network with [466, 167, 234] neurons in each layer. This architecture is retained throughout this work. 

\subsubsection{Number of latent variables}
Finally, we study the effect of increasing the number of latent variables in the autoencoder, fixing the three-layer network with [466, 167, 234] neurons in each layer (and optimized in the previous section). We vary the number of latent variables by changing the width of the middle layer of the optimal autoencoder network with the architecture described in the previous subsection. 

\begin{figure}[htbp]
    \centering
    \includegraphics[scale=0.8]{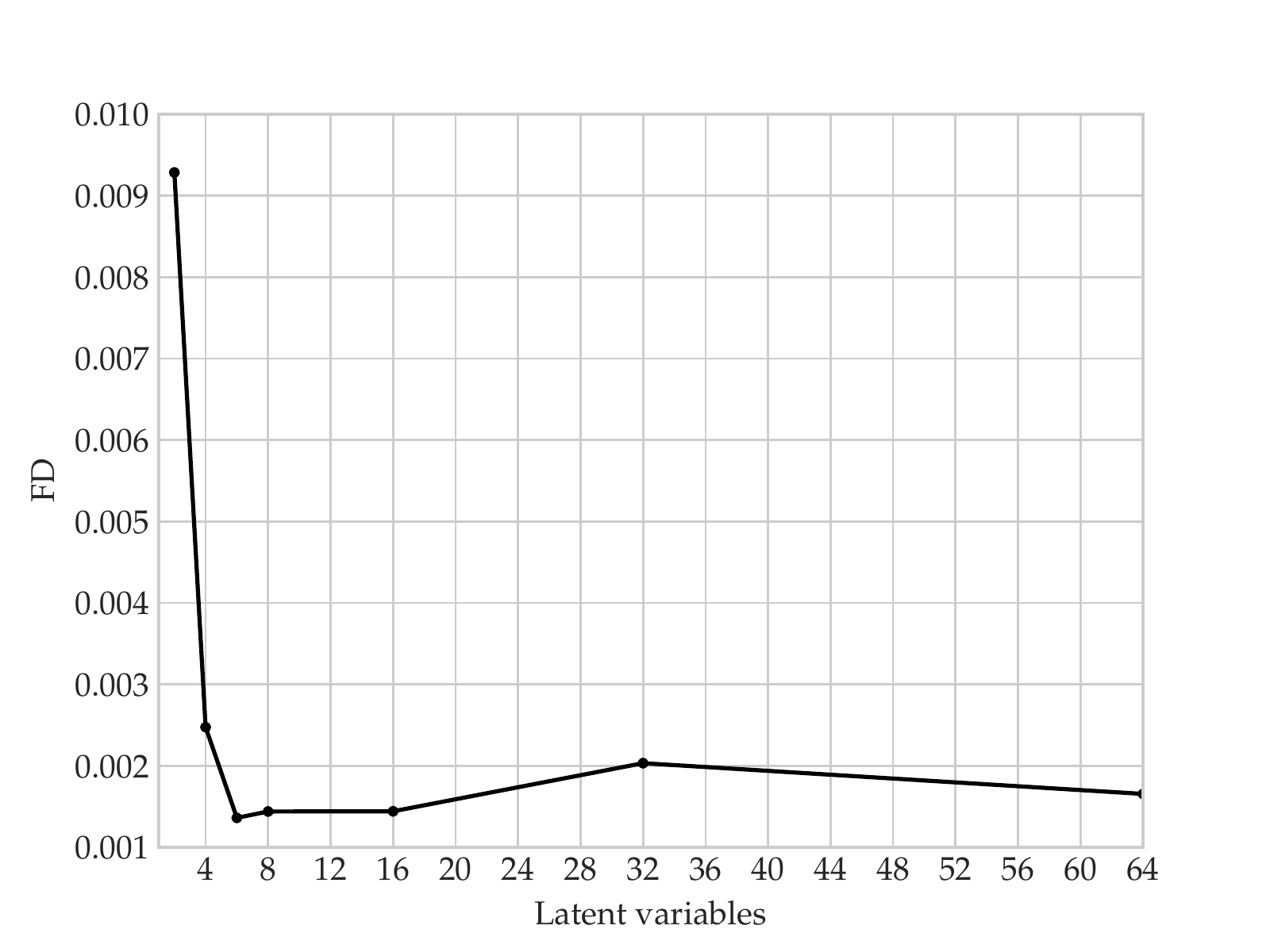}
    \caption{Fréchet distance against number of latent variables for a three-layer network with [466, 167, 234] neurons in each layer.}
    \label{fig:FDvsn}
\end{figure}

Figure \ref{fig:FDvsn} shows the Fréchet distance variations when changing the number of latent variables. We observe a sharp increase in the Fréchet distance for $n<4$ and little improvement for $n>4$. We keep $n=4$ latent variables. Note that the latent variables are the variables for shape optimization and thus using a small number leads to fast optimizations.  

\subsection{Optimization algorithm}
To efficiently explore the global parameter space, an evolutionary algorithm is considered for optimization. In particular, the particle swarm optimization (PSO) algorithm is used for optimization. PSO is a population-based evolutionary heuristic algorithm proposed by Eberhart and Kennedy \cite{kennedy1995particle}. It basically describes a simplified social model that stems from swarming theory, which mimics the swarm of animal adapting to its surrounding environment. Similar to other evolutionary algorithms \cite{zhou2011multiobjective}, PSO does not require any gradient information (gradient-free); therefore, it is very useful for global optimization. So far, PSO has been applied to different aerodynamic optimization problems \cite{venter2004multidisciplinary,wickramasinghe2010designing,xia2017particle}.

In the PSO algorithm, the particles are initially populated at random positions, which represent a candidate optimization solution. The fitness function is defined to represent the knowledge gained for each particle, which defines a criterion that determines the proximity of a particle to the optimal solution. The basic algorithm of PSO is detailed in \cite{kennedy1995particle}, which includes four steps as follows:
\begin{enumerate}
\item Initialize the particle swarm with random positions and random initial velocities;
\item Update the velocity for each particle, using the history of evolution and the knowledge (obtained from the fitness function) gained by the swarm;
\item Update the position of each particle based on the updated velocity and the previous position;
\item Go to Step 2 and repeat until convergence.
\end{enumerate}
At each iteration, the position and velocity of the $i$th particle are updated as follows:

\begin{equation}
\label{Eq:PSO-vel}
    \boldsymbol{V}^{g+1}_i = \omega (g) \times \boldsymbol{V}^{g}_i + c_1 r_1 \times (\boldsymbol{Pbest}^{g}_i - \boldsymbol{X}^{g}_i) + c_2 r_2 \times (\boldsymbol{Gbest}^{g}_i - \boldsymbol{X}^{g}_i),
\end{equation}
\begin{equation}
    \boldsymbol{X}^{g+1}_i = \boldsymbol{X}^{g}_i + \boldsymbol{V}^{g+1}_i ,
\end{equation}
\begin{equation}
    \omega(g+1) = \omega_2 - (\omega_2 - \omega_1) \times g / G ,
\end{equation}
where $g$ is the index of the present generation (iteration), $G$ is the maximum generation (iteration) number. $\boldsymbol{X}^{g+1}_i$ and $\boldsymbol{V}^{g+1}_i$ denote the position and velocity of the $i$th particle in the $g+1$th iteration, respectively; $\boldsymbol{Pbest}^{g}_i$ and $\boldsymbol{Gbest}^{g}_i$ denote the best local and global positions, respectively. To achieve an effective local search, a linear time-varying weight $\omega(g+1)$ decreases linearly with an increasing number of iterations \cite{shi1998modified}. $c_1$ and $c_2$ are two acceleration constants, $r_1$ and $r_2$ are random numbers from $0$ to $1$ and $\omega_1$ and $\omega_2$ are the lower and upper bounds of the weight. The velocity is updated according to the theory of swarm intelligence, as described in Equation \ref{Eq:PSO-vel}. The first term reflects the memory effects of the velocity. The second term refers to the current knowledge in the swarm, whereas the third term refers to the group knowledge in the swarm. In practice, the maximal and minimal values of the particle positions and velocities should be limited to avoid crossing the boundaries of the search space. The selection of PSO parameters has been well discussed in \cite{pedersen2010good,juneja2016particle,wang2018particle}. 

\subsection{Problem formulation: optimization objective and constraints}
In this study, we are interested in geometric optimization for noise minimization. Therefore, the Overall A-weighted Sound Pressure Level (OASPL) is selected as the objective function. The OASPL is used to weight the spectrum energy in the frequency domain with the sensitivity of human hearing, resulting in optimized geometries suitable for wind turbines located near urban areas. Since the optimized airfoil is expected to work on multiple operating conditions, multi-objective optimization is formulated where the overall OASPL for various angles of attack (AOAs) is minimized. Meanwhile, to maintain reasonable aerodynamic performance and reasonable geometries, several constraints should be satisfied. Therefore, the following objective and constraints are selected:

\begin{equation}
\label{eq:objective}
    \min_{b / z} \sum_{i=1}^s OASPL(AOA_i),
\end{equation}

\begin{align}
    \text{subject to } & C_{l,i} \geq C_{l0,i}\;\;;\;\; C_{d,i} \leq C_{d0,i}, \nonumber\\
    & Thickness \leq Thickness_0 , \nonumber\\
    & \min (y_{upper}(x)-y_{lower}(x)) > 0\;\;;\;\;  0<x<1, \nonumber\\
    & Thickness_{\text{Leading}} \geq Thickness_\text{Trailing}.\nonumber
\end{align}
where $b$ and $z$ refer to the CST parameters or the latent variable of the autoencoder. The sum of OASPL at the $s$ angle of attack is used to evaluate the objective function. The parameter with subscript $0$ refers to the baseline geometry, where the NACA0012 airfoil is chosen for all angles of attack. To maintain reasonable aerodynamic performance, we do not expect a drop in lift or an increase in drag, compared to the baseline case. Therefore, the first group of constraints is imposed: $C_{l,i} \geq C_{l0,i}, C_{d,i} \leq C_{d0,i}$, where $C_{l0}$ and $C_{d0}$ refer to the lift and drag coefficients of the baseline geometry, respectively. The constraint applies to all AOAs considered during optimization. All of them are hard constraints that must be satisfied during the optimization process; therefore, a suitable constraint handling technique is needed and will be explained hereafter. 

Apart from aerodynamic considerations, the constraints ensure the generation of realistic airfoil geometries. From previous studies on aeroacoustic optimization \cite{brianaerodynamic,zhou2015discrete,monfaredi2021unsteady}, optimal aeroacoustic geometry generally does not produce improved aerodynamic performance. Therefore, to improve the optimization efficiency, geometrical considerations are introduced as constraints. Although a thinner airfoil is expected to be more silent, the second constraint ($Thickness \leq Thickness_0 $) limits the thickness of the airfoil to avoid being thicker than the benchmark airfoil. This is required to ensure structural capabilities/resistance in the final geometry \cite{ricks2020cfd}. The third constraint $\min (y_{upper}(x)-y_{lower}(x)) > 0\;;\; 0<x<1,$ ensures that the upper and lower surfaces do not cross each other (e.g., negative volume shapes). To satisfy this condition, the difference between the upper layer and the lower layer, given the same $x$ coordinate, should always be positive. Moreover, the last constraint $Thickness_{\text{Leading}} \geq Thickness_{\text{Trailing}} $ demands that the first half of the airfoil is thicker than the second half, to ensure a reasonable aerodynamic configuration. The two thickness parameters are defined as follows:

\begin{eqnarray}
   Thickness_{\text{Leading}} = \text{max}(y_{upper}(x)- y_{lower}(x))\;\;;\;\; 0 \leq x \leq 0.5, \\
%
   Thickness_{\text{Trailing}} = \text{max}(y_{upper}(x) - y_{lower}(x))\;\;;\;\; 0.5 < x \leq 1.\nonumber
\end{eqnarray}

\begin{figure}[htbp]
    \centering
    \includegraphics[width=250pt]{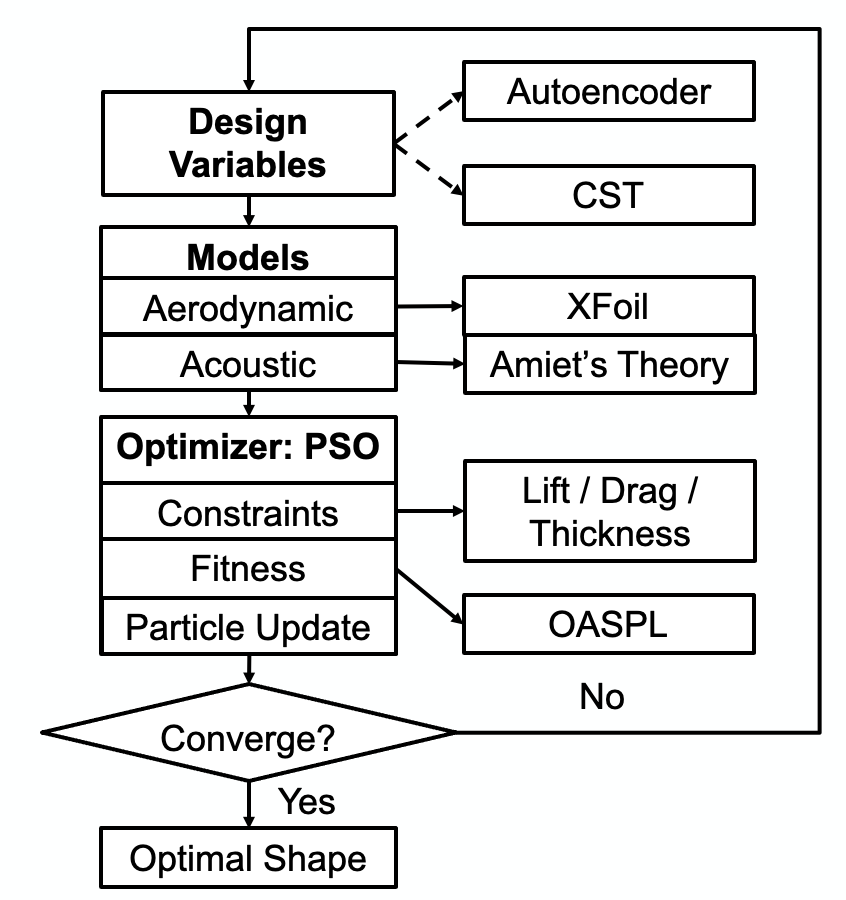}
    \caption{Aeroacoustic optimization framework.}
    \label{fig:framework}
\end{figure}

The optimization framework is illustrated in Fig. \ref{fig:framework}. As shown in the figure, the entire optimization framework involves three main components: 1) design variables (geometrical parametrization); 2) aerodynamic and aeroacoustic models; 3) optimizer (optimization algorithm and constraint handling). In the current study, the second and third components are fixed, while two different methods for geometrical parametrization have been considered (CST and autoencoder) to show the potential of machine learning to enhance aeroacoustic shape optimization.

It should be noted that the optimization results depend heavily on the problem formulation \cite{brianaerodynamic}. Therefore, if other constraints are used, the results can be different. In the present work, the aim is not to propose a new acoustic geometry but rather to show the capability of autoencoders to enhance the optimization problem. To complete the optimization algorithm, robust constraint handling techniques should be considered \cite{biswas2020multi,kuk2021empirical}. Different methods for constraint handling have been summarized by Kuk et al. \cite{kuk2021empirical}, such as discarding / modifying unfeasible solutions and transforming objection functions, etc. Here, we choose to replace unfeasible solutions by randomly generating new solutions \cite{roy2013multi}, which is beneficial for global searches. At each PSO iteration, the particle is first evaluated based on the constraints, and is discarded if the constraints are not satisfied. The discarded particle will be replaced by a new particle randomly generated in the sampling space.

Finally, the range of shape parameters, including the CST shape parameters or latent variable of the autoencoder, should be predetermined before the optimization starts. For CST parametrization, we have observed that if the parameter ranges are not properly chosen, XFOIL can easily stagnate into an infinite loop that hinders the convergence of the optimization. This is due to the fact that CST does not always produce reasonable geometries. Therefore, we chose to perturb the CST parameters of a baseline airfoil (the NACA0012 airfoil), and the CST parameters are allowed to vary between $\pm 30\%$ of the baseline parameters. In contrast, since the autoencoder is trained from a database of real airfoils, it generates new geometry that follows the training database. Therefore, the entire optimization process is easier to converge; thus a wider range of geometrical parameters (latent variables) can be selected. The optimization process converges when the predefined number of iterations / generations is reached.

\section{Results and discussion}
\label{sec:results}
The initial case to start the optimization procedure is a widely studied airfoil in the literature, i.e., a NACA~0012 airfoil, in similar conditions as the validation case detailed in Section \ref{sec:validation}, with the difference that it is assumed that the airfoil has a $1~m$ chord and span. The far-field observer is located at $1~m$ from the airfoil trailing edge in the midspan, and the wall pressure spectrum is calculated at $x/c~=~0.99$. The boundary layer transition is fixed at $x/c~=~0.06$ on both sides. The inflow velocity is assumed to be $56~m/s$, which corresponds to a Reynolds number
3.7~$\times$~10\textsuperscript{6} and Mach number 0.16, typical for small wind turbine blades. The frequency resolution is 1~Hz and the far-field noise $S_\mathrm{pp}$ is calculated from 100~Hz to 10~kHz. Since the optimized airfoil is expected to operate under multiple operating conditions, the objective function in Eq. \ref{eq:objective} includes the sum of OASPL from three typical AOAs: 0\textdegree, 2\textdegree, and 4\textdegree. For each OAO, the OASPLs are calculated first by applying the A-weighted filter~\cite{IEC_61672} in the entire frequency spectrum, then calculating the SPL, using Equation~\ref{Eq:SPL}, and finally by integrating in the entire frequency range (which is the output of our acoustic simulator).

\subsection{Classical optimization with CST parametrization}
In the first group of optimization, the CST parametrization is used. A NACA0012 airfoil is used for baseline geometry, which has been widely used for case validation. Six CST parameters have been chosen for each surface, to sufficiently represent the geometry. The CST parameters for NACA0012 airfoil are:

\begin{eqnarray}
    Upper:& [0.170374, 0.160207, 0.143643, 0.166426, 0.110476, 0.179433],\\
%
    Lower:& [-0.170374, -0.160207, -0.143643, -0.166426, -0.110476, -0.179433].\nonumber
\end{eqnarray}
Note that because of the symmetrical configuration, the CST parameters for the lower surface are the negative counterparts of the upper surface. From the baseline parameters, we allow $\pm 30\%$ to change the CST parameters. In the PSO optimization, the number of iterations / generation is set to 80, while a group of 40 particles is initialized. 

\begin{figure}[htbp]
    \centering
    \includegraphics[width=250pt]{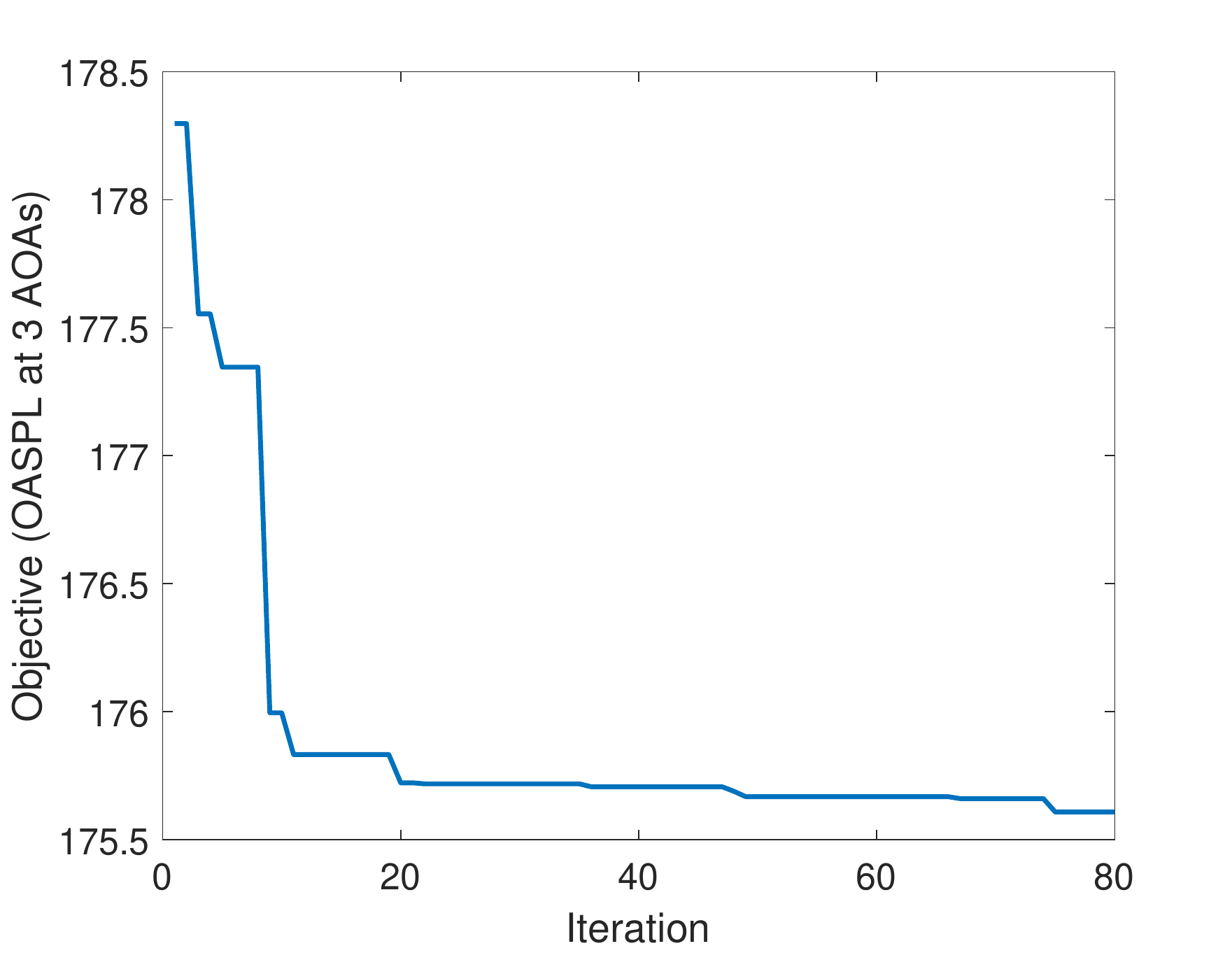}
    \caption{Convergence history in the optimization process.}
    \label{fig:history_CST}
\end{figure}

\begin{figure}[htbp]
    \centering
    \includegraphics[width=250pt]{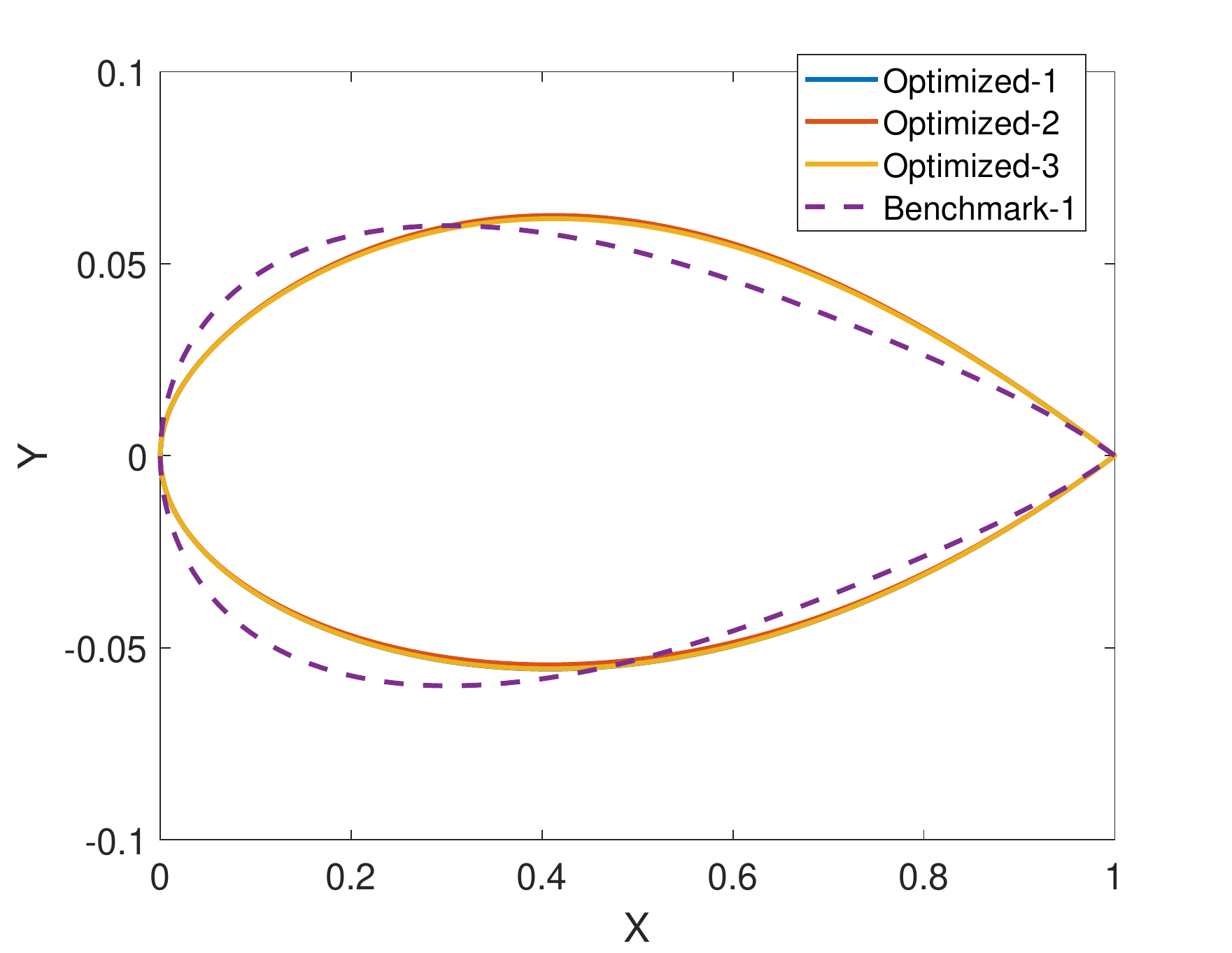}
    \caption{Optimized geometry versus the baseline geometry (NACA0012 airfoil) based on CST parametrization. Benchmark geometry is generated from the mean of CST parameters, which is the NACA0012 airfoil in this work.}
    \label{fig:airfoil_CST}
\end{figure}

Fig. \ref{fig:history_CST} shows a typical convergence plot in the optimization process. As the PSO iterations progress, the objective function drops first rapidly and then converges to the minimal values. From the first to the final iteration, about 3dBA noise is reduced for all AOAs. We run the optimization several times, resulting in very similar final geometries. Three optimized shapes are shown in Fig. \ref{fig:airfoil_CST}, which are very similar to each other. Compared to the benchmark profile, the optimized ones are thinner near the leading edge and thicker near the trailing edge. The predicted aerodynamic and acoustic parameters will be compared and discussed in the next section, as shown in Table \ref{table1},  Table \ref{table2} and Table \ref{table3}

\subsection{Improved optimization with autoencoder-based parametrization}
\begin{figure}[htbp]
    \centering
    \includegraphics[width=250pt]{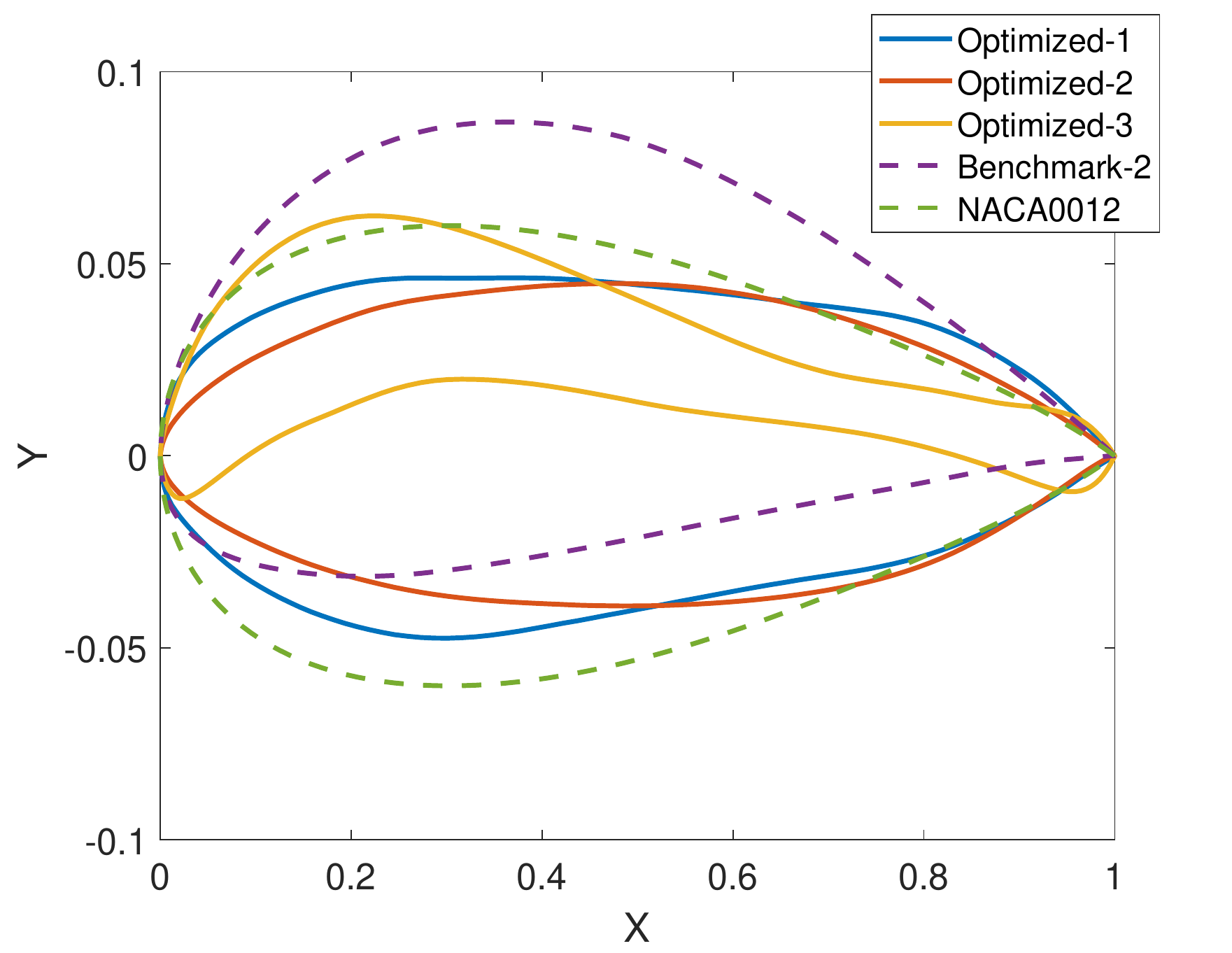}
    \caption{Optimized geometry versus the NACA0012 airfoil based on autoencoder parametrization. Benchmark geometry is generated from the mean of latent variables of autoencoder, which are all set to zero in this work.}
    \label{fig:airfoil_autoencoder}
\end{figure}

In this section, the same optimization problem is solved with an autoencoder used for shape parametrization. We train the autoencoder with 4 latent variables, which are sufficient to represent the two-dimensional aerodynamic configurations. The same PSO parameters are used during the optimization. As detailed in Section \ref{sec:autoencoder}, the autoencoder generates reasonably good aerodynamic configurations across a wide range of latent variables. Therefore, the range of all four latent variables is set to $[-3,3]$. All PSO parameters are set the same as those used for CST-based optimization. In contrast to when using CST-based optimization, the autoencoder provides multiple final geometries. Three optimum geometries are depicted in Fig. \ref{fig:airfoil_autoencoder}. When checking the aerodynamic coefficients and the OASPL, see Table \ref{table1}, we observe that all geometries have similar values for the objective function, but the profiles are now different. In this case, the benchmark geometry is generated by setting all the latent variables to zero (and does not correspond to the NACA0012). 

\begin{table}[htbp!]
	\vspace{20pt}
	\centering
	\small
	\begin{tabular}{p{2.5cm}p{2cm}p{3.5cm}p{4cm}p{0.9cm}p{0.9cm}p{1.1cm}p{0.9cm}p{0.9cm}p{1.1cm}}
  \hline
 Method & {Objective } & $\%$ Noise Reduction & dBA Noise Reduction \\
 & OASPL & w.r.t NACA0012 & w.r.t NACA0012 \\
  \hline
  NACA0012 & 178.9352 & $0$ & $0$ \\
  CST-1 & 175.6013 & $1.86$ & 1.1113 \\
  CST-2 & 175.6054 & $1.86$ & 1.1099 \\
  CST-3 & 175.6087 & $1.86$ & 1.1083 \\
  Autoencoder-1 & 173.6856 & $2.93$ & 1.7499 \\
  Autoencoder-2 & 175.6192 & $1.85$ & 1.1053 \\
  Autoencoder-3 & 176.0466 & $1.61$ & 0.9629 \\
  \hline
 \end{tabular}
	\caption{Objective function for different optima and overall noise reduction when compared to a NACA0012 airfoil.}
	\label{table1}
\end{table}

\begin{figure*}[htbp]
\begin{subfigure}{.45\textwidth}
		\includegraphics[width=180pt]{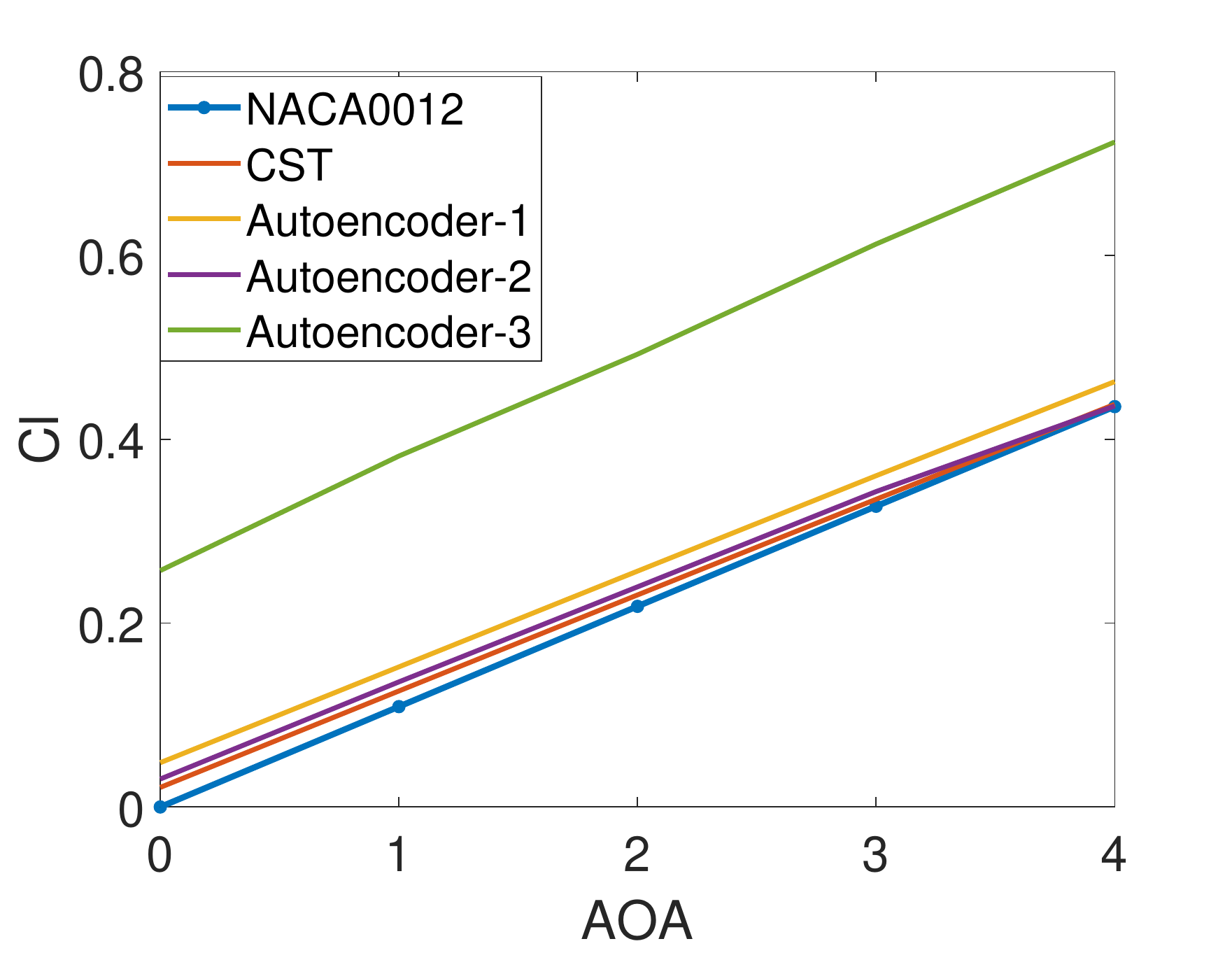}
		\caption{}
	\end{subfigure}
	\begin{subfigure}{.45\textwidth}
		\includegraphics[width=180pt]{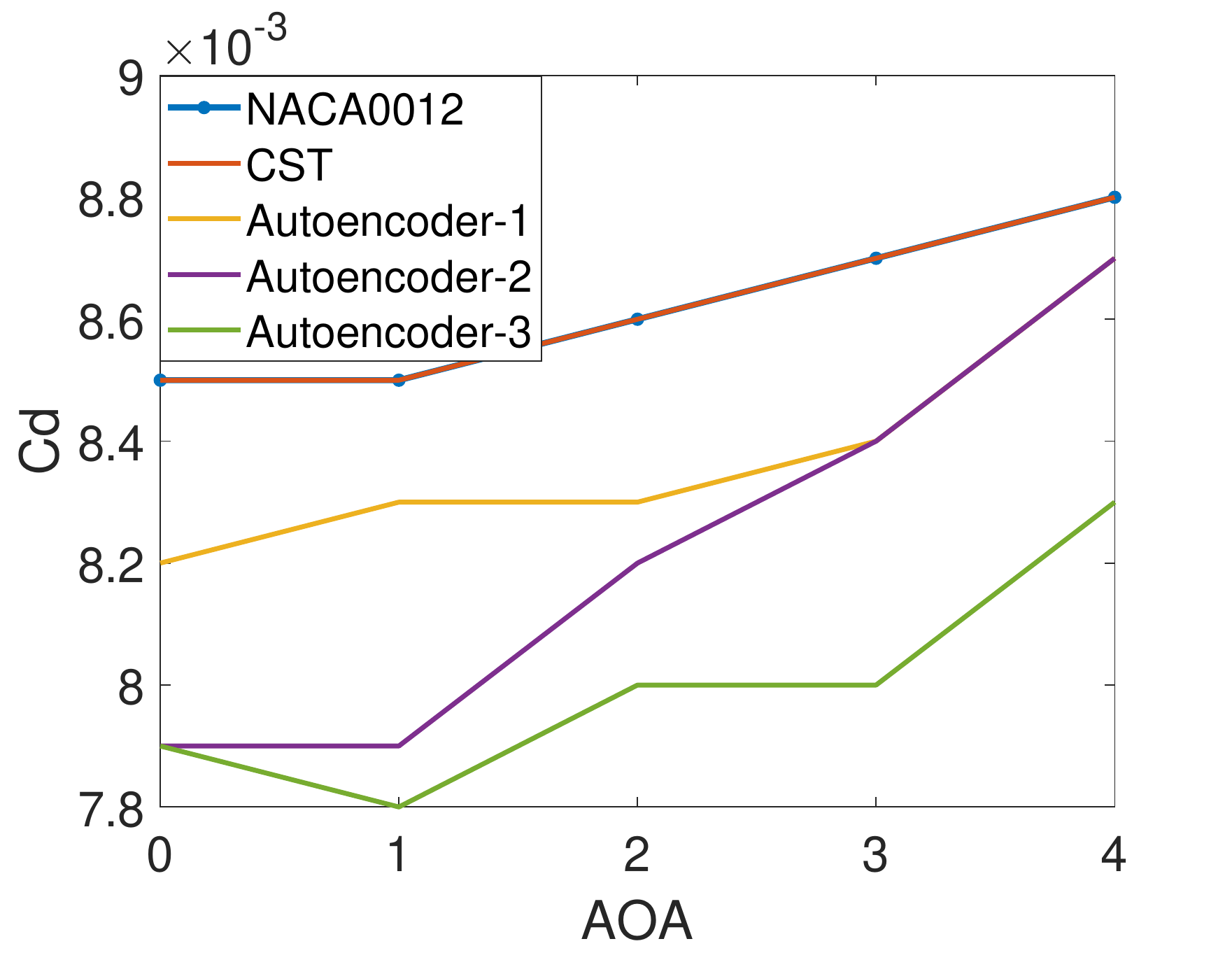}
		\caption{}
	\end{subfigure}
    \begin{subfigure}{.45\textwidth}
		\includegraphics[width=180pt]{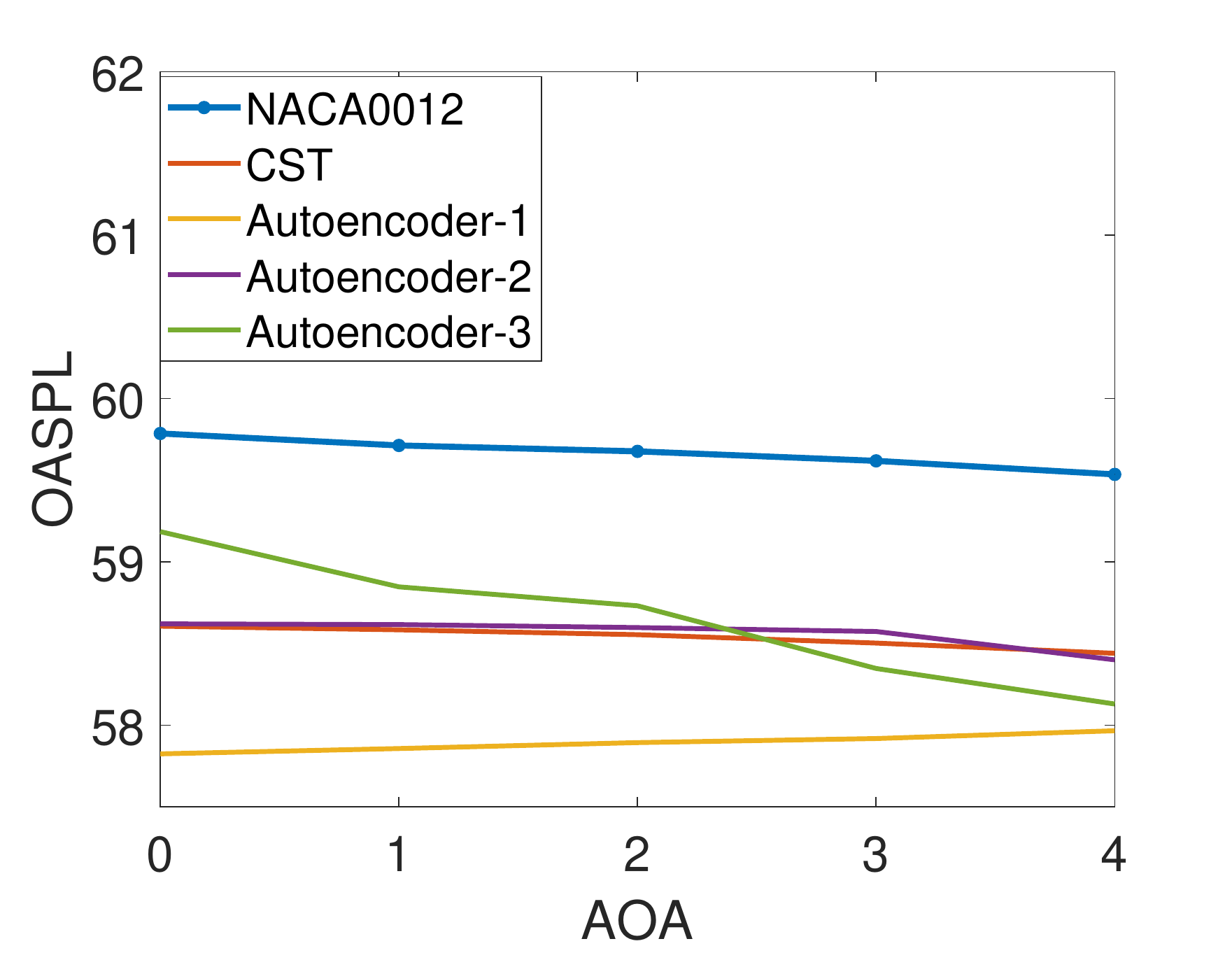}
		\caption{}
	\end{subfigure}
	\centering
	\caption{Comparison of aerodynamic and acoustic parameters for optimized airfoils. a) Cl. b) Cd. c) OASPL.}
	\label{fig:Aero}
\end{figure*}

\begin{figure*}[htbp]
\begin{subfigure}{.45\textwidth}
		\includegraphics[width=180pt]{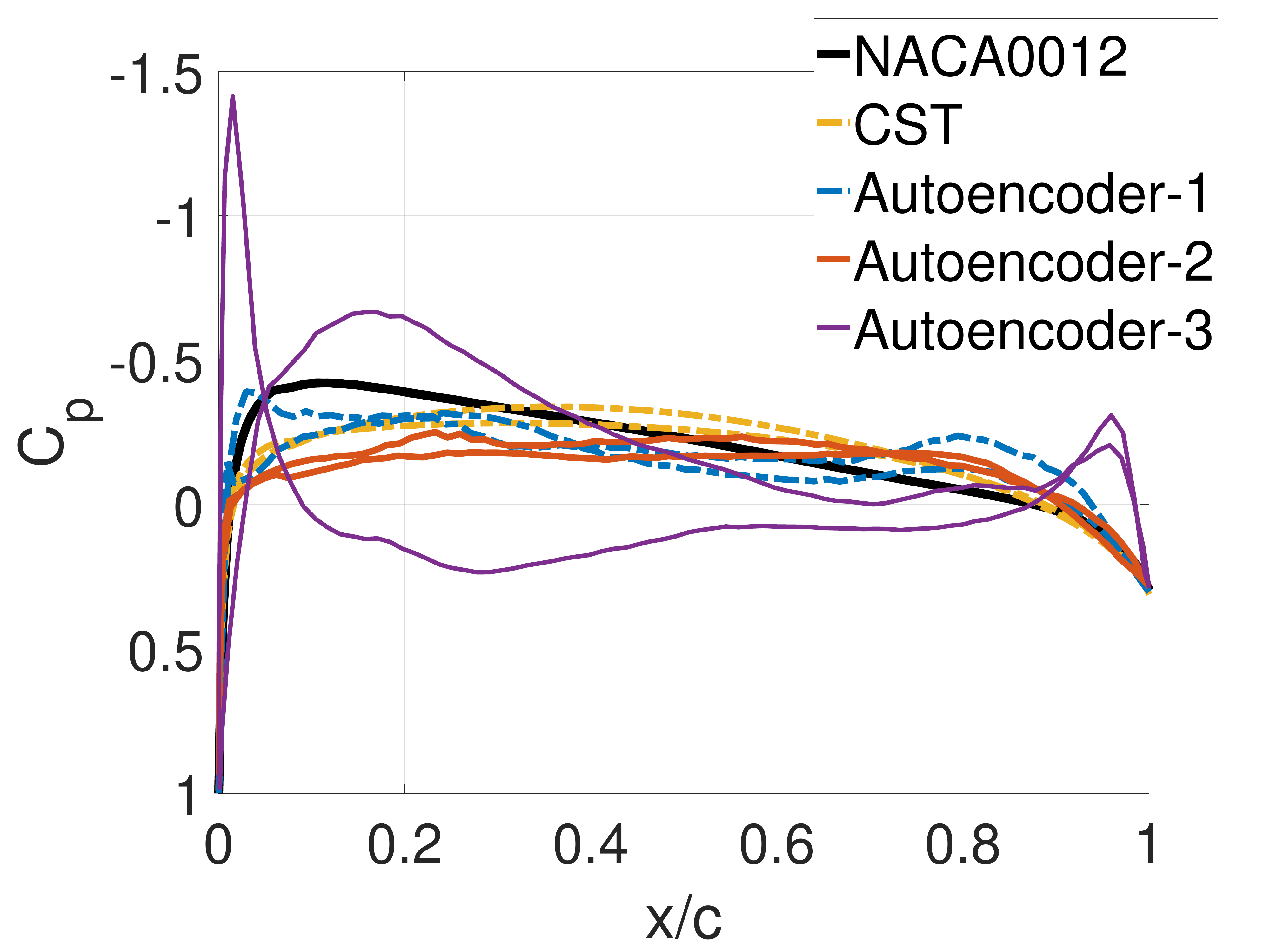}
		\caption{}
	\end{subfigure}
	\begin{subfigure}{.45\textwidth}
		\includegraphics[width=180pt]{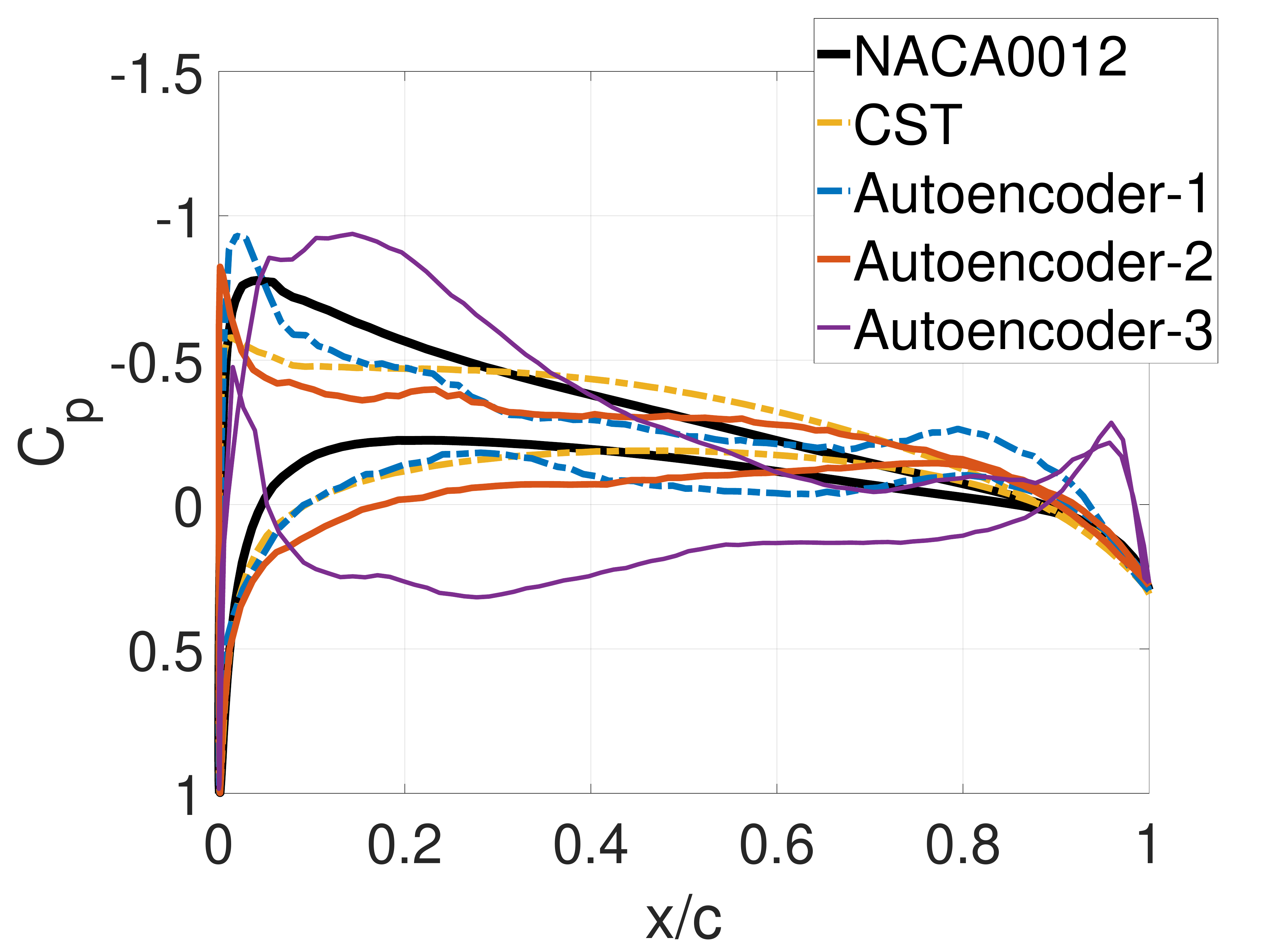}
		\caption{}
	\end{subfigure}
    \begin{subfigure}{.45\textwidth}
		\includegraphics[width=180pt]{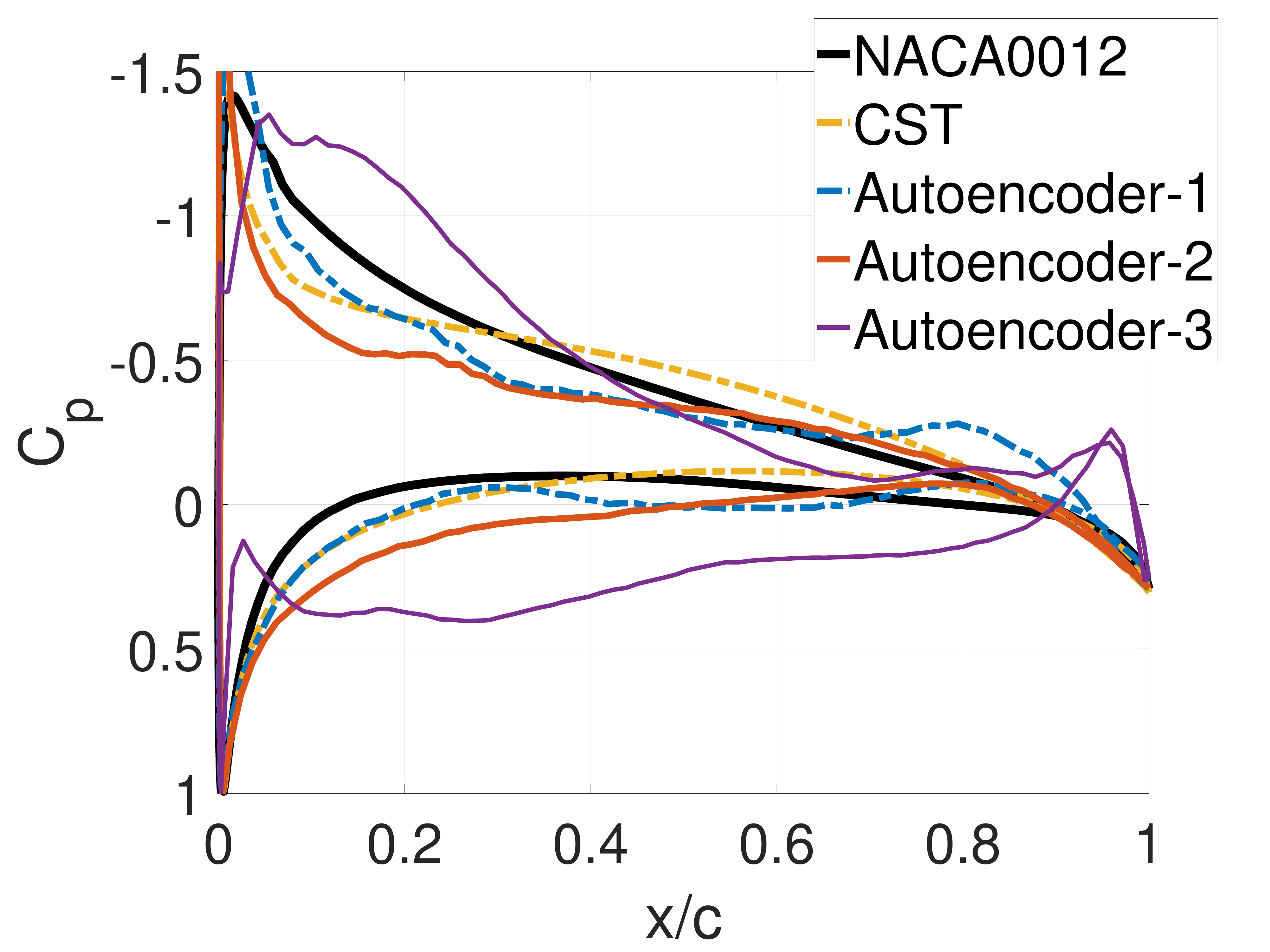}
		\caption{}
	\end{subfigure}
	\centering
	\caption{Comparison of surface pressure coefficient for optimized airfoils. a) AOA = 0\textdegree. b) AOA = 2\textdegree. c) AOA = 4\textdegree.}
	\label{fig:Cp}
\end{figure*}

The values of all objective functions, including both CST-based and autoencoder-based optimization, are summarized in Table \ref{table1}. As shown in the table, both types of parametrization lead to very similar objective values, but the autoencoder finds more silent airfoils. The aerodynamic and acoustic behaviors are compared in Table \ref{table2} (for all AOAs used for optimization, including 0\textdegree, 2\textdegree and 4\textdegree) and Table \ref{table3} (for other AOAs including 1\textdegree  and 3\textdegree). These results are plotted in Fig. \ref{fig:Aero}. Although having similar OASPL, the aerodynamic properties from autoenconder-based optimization are better than those from CST-based optimization. Furthermore, the optimal airfoil in terms of OASPL (Autoencoder-1) reduces around $2.97\%$ for all angles of attack (the mean value, corresponding to 1.75 dBA), compared to the baseline NACA0012 airfoil. This indicates that the autoencoder can fully explore the parameter space to generate optimal airfoils. For completeness, the pressure coefficients of all configurations are compared in Fig. \ref{fig:Cp}.

\begin{table}[htbp!]
	\vspace{20pt}
	\centering
	\scriptsize
	\begin{tabular}{p{2cm}p{0.8cm}p{0.8cm}p{0.8cm}p{0.8cm}p{0.8cm}p{0.8cm}p{0.8cm}p{0.8cm}p{0.8cm}}
  \hline
 Shape & \multicolumn{3}{c}{AOA=0} & \multicolumn{3}{c}{AOA=2} & \multicolumn{3}{c}{AOA=4} \\
   & $C_l$ & $C_d$ & $OASPL$ & $C_l$ & $C_d$ & $OASPL$ & $C_l$ & $C_d$ & $OASPL$ \\
  \hline
  NACA0012 & 0 & 0.0085 & 59.7859 & 0.2183 & 0.0086 & 59.6765 & 0.4356 & 0.0088 & 59.5355 \\
  CST & 0.0213 & 0.0085 & 58.6063 & 0.2307 & 0.0086 & 58.5545 & 0.4381 & 0.0088 & 58.4405 \\
  Autoencoder-1 & 0.0482 & 0.0082 & 57.8251 & 0.2566 & 0.0083 & 57.8940 & 0.4628 & 0.0087 & 57.9665 \\
  Autoencoder-2 & 0.0304& 0.0079 & 58.6207 & 0.2395 &  0.0082 &  58.5979 & 0.4363 & 0.0087 & 58.4006 \\
  Autoencoder-3 & 0.2572 & 0.0079 & 59.1851 & 0.4926 & 0.0080 & 58.7313 & 0.7235 & 0.0083 & 58.1302 \\
  \hline
 \end{tabular}
	\caption{Aerodynamic and aeroacoustic parameters for different geometries at AOAs used for optimization (0\textdegree, 2\textdegree and 4\textdegree). OASPL values are presented in dBA.}
	\label{table2}
\end{table}

\begin{table}[htbp!]
	\vspace{20pt}
	\centering
	\begin{tabular}{p{2.7cm}p{1cm}p{1cm}p{1.3cm}p{1cm}p{1cm}p{1.3cm}p{1cm}p{1cm}p{1.3cm}}
  \hline
 Shape & \multicolumn{3}{c}{AOA=1} & \multicolumn{3}{c}{AOA=3} \\
   & $C_l$ & $C_d$ & $OASPL$ & $C_l$ & $C_d$ & $OASPL$\\
  \hline
  NACA0012 & 0.1092 & 0.0085 & 59.7126 & 0.3272 & 0.0087 & 59.6183 \\
  CST & 0.1262 & 0.0085 & 58.5842 &  0.3348 & 0.0087 & 58.5031 \\
  Autoencoder-1 & 0.1524 & 0.0083 & 57.8576 & 0.3604 & 0.0084 & 57.9188 \\
  Autoencoder-2 & 0.1361 & 0.0079 & 58.6163 & 0.3432 & 0.0084 & 58.5739 \\
  Autoencoder-3 & 0.3818 & 0.0078 & 58.8472 & 0.6125 & 0.0080 & 58.3477\\
  \hline
 \end{tabular}
	\caption{Aerodynamic and aeroacoustic parameters for different geometries (non-optimized operation conditions, 1\textdegree and 3\textdegree). OASPL values are presented in dBA.}
	\label{table3}
\end{table}

\begin{figure*}[htbp]
    \begin{subfigure}{.45\textwidth}
		\includegraphics[width=180pt]{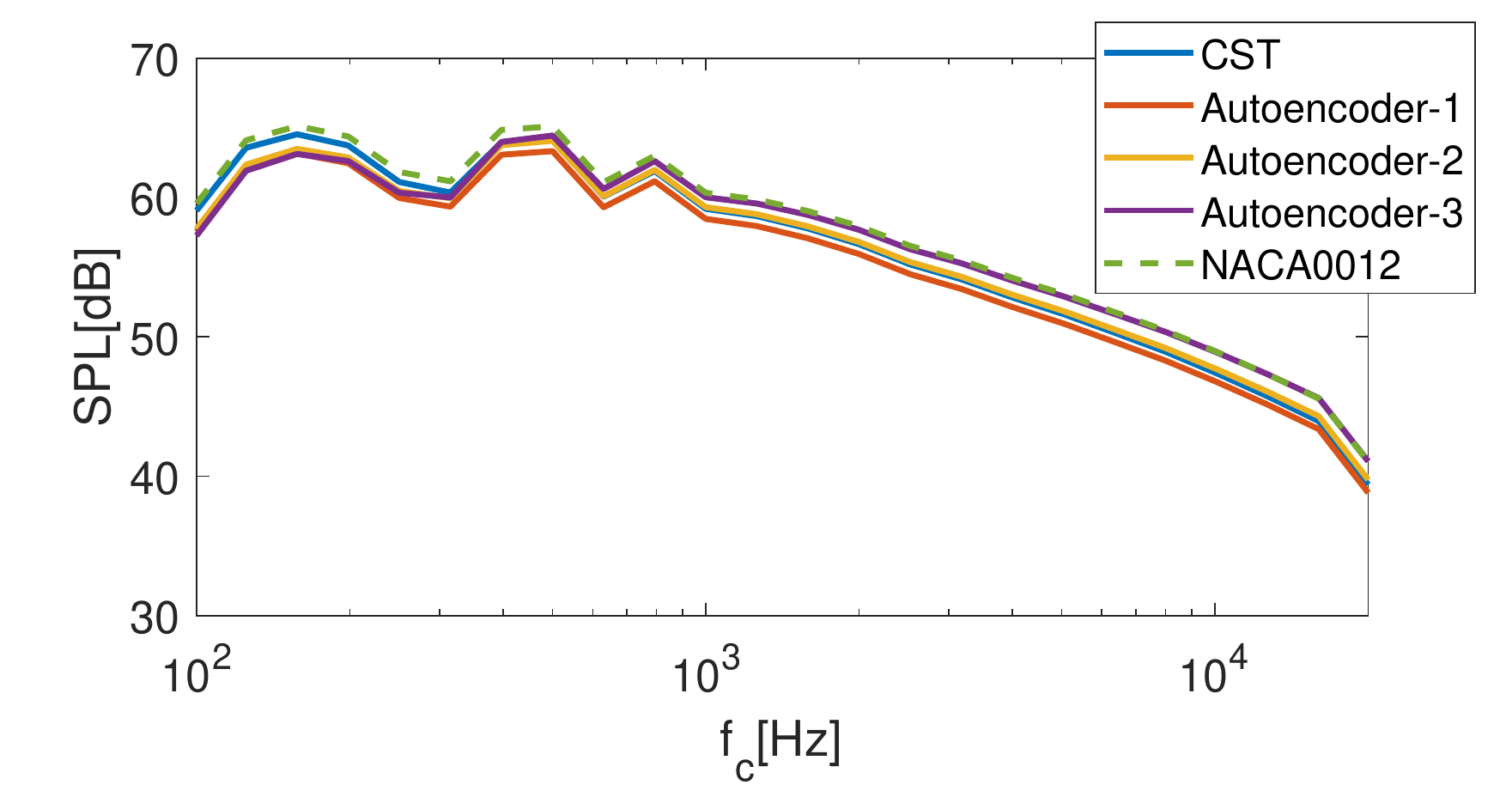}
		\caption{}
	\end{subfigure}
	\begin{subfigure}{.45\textwidth}
		\includegraphics[width=180pt]{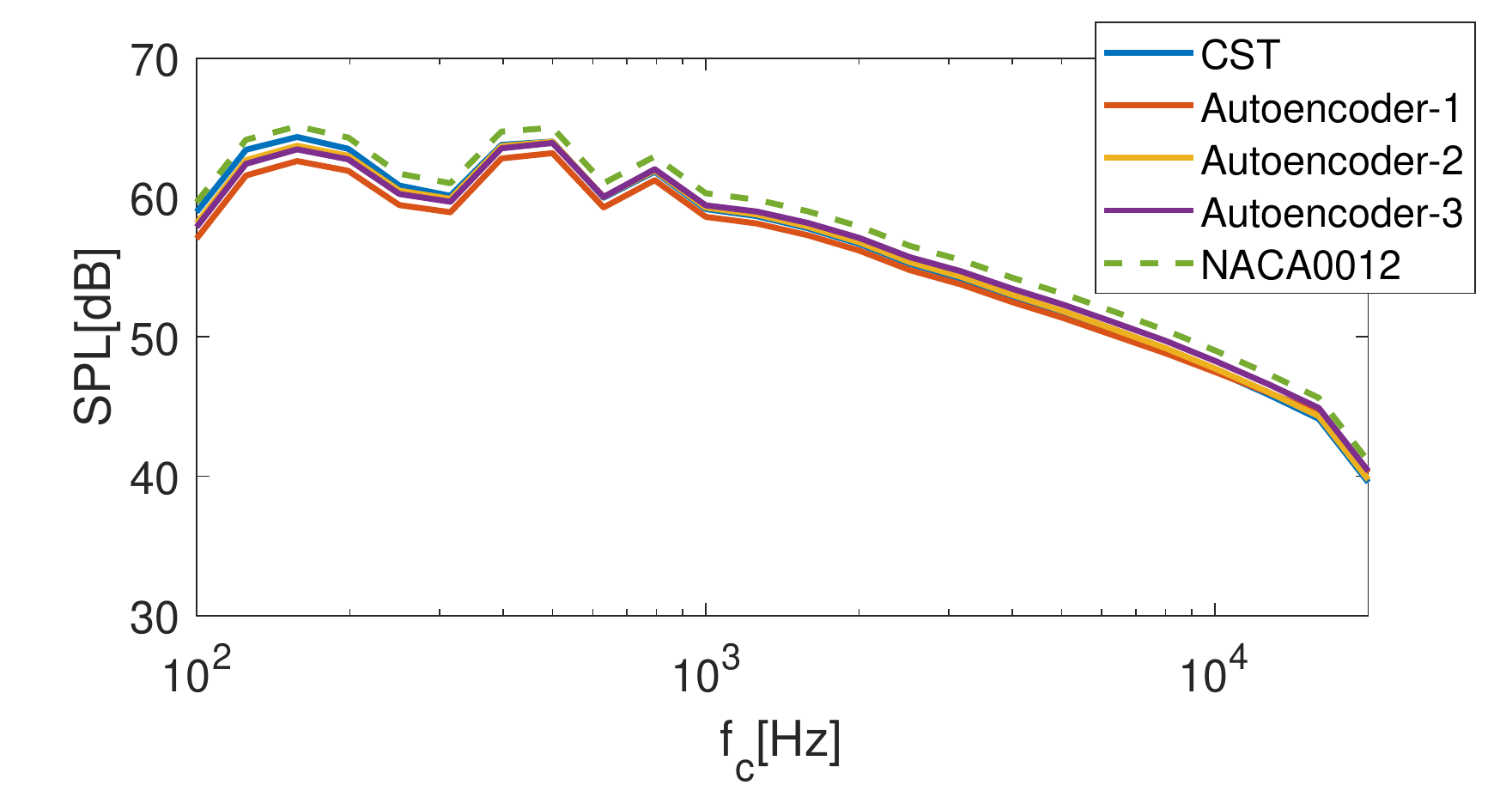}
		\caption{}
	\end{subfigure}
    \begin{subfigure}{.45\textwidth}
		\includegraphics[width=180pt]{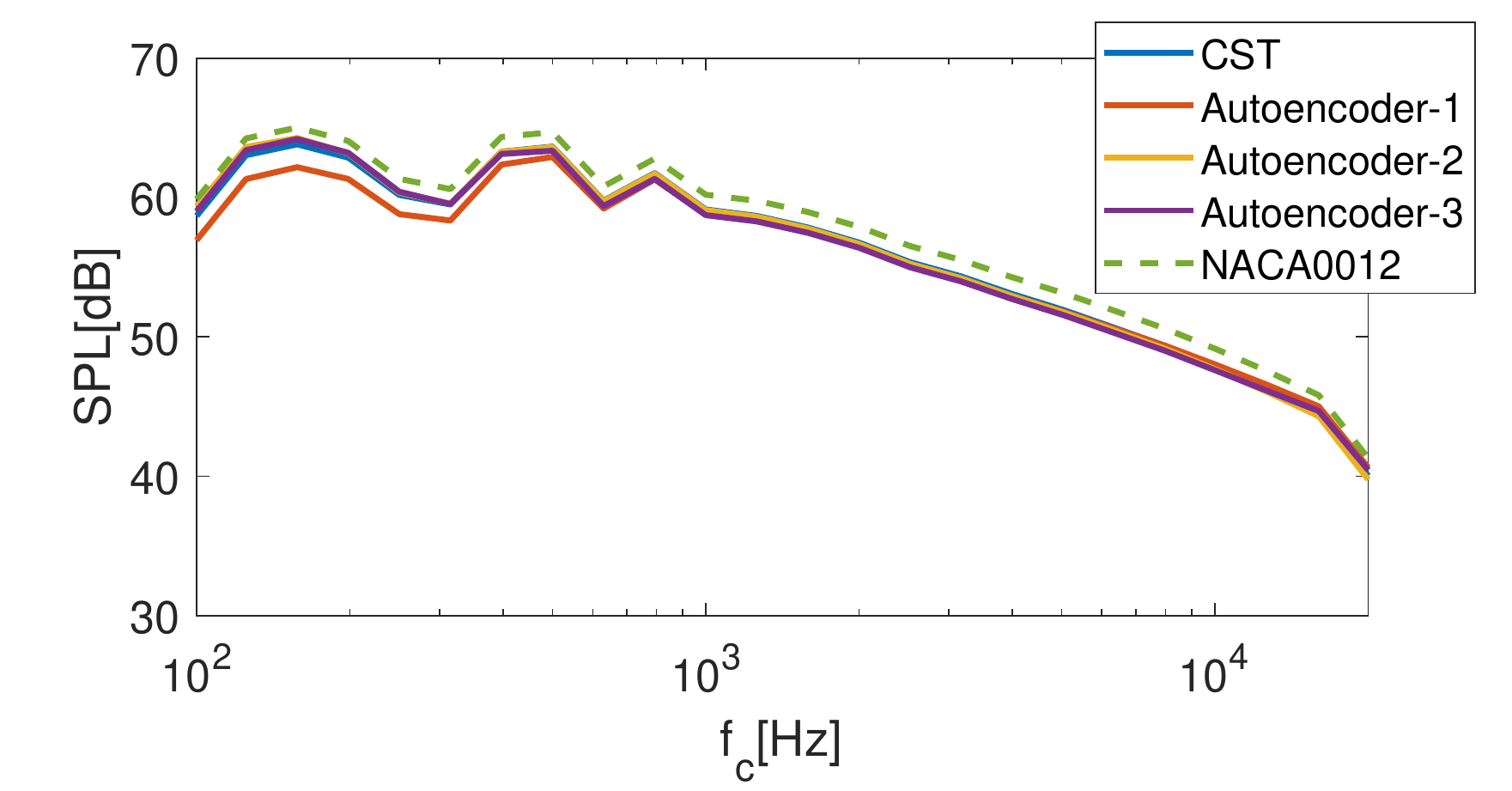}
		\caption{}
	\end{subfigure}
	\centering
	\caption{Far field noise comparison for optimized airfoils. Sound pressure levels are presented in $1/3$ octave band. a) AOA = 0\textdegree. b) AOA = 2\textdegree. c) AOA = 4\textdegree.}
	\label{fig:Noise}
\end{figure*}

To compare the noise reduction for optimized configurations, Fig. \ref{fig:Noise} shows the far-field noise spectrum. It is clear that, although the integrated value (OASPL) is used as the objective, the noise across all frequency ranges is reduced. Compared with the baseline NACA0012, the optimized airfoils reduce noise at all frequencies and all angles of attack considered. At a small angle of attack (e.g., AOA = 0\textdegree), optimized airfoils based on autoencoders show lower noise levels at small frequencies, compared to optimized airfoils based on CST. On the contrary, the optimized airfoil based on the third autoencoder shows louder noise at high frequencies, similar to the baseline airfoil. As the angle of attack is increased, CST, Autoencoder-2 and Autoencoder-3 airfoils behave similarly, producing less noise compared to the NACA0012 airfoil. Additionally, for all operating conditions, the Autoencoder-1 airfoil gives minimal overall noise for all frequencies, which is consistent with the fact that this airfoil has minimal objective function. This indicates that OASPL is a good indicator used for the acoustic optimization of airfoils and that variational autoencoders are a promising tool for airfoil optimization. 

\section{Conclusions}
\label{sec:conclusion}
This work develops an aeroacoustic airfoil shape optimization framework to reduce the trailing edge noise produced by an attached turbulent boundary layer. We establish the numerical framework by using Amiet's theory coupled with the TNO-Blake model for acoustics and the XFOIL solver for aerodynamics. We use PSO in the optimization process to ensure an efficient global search. We define the objective function of the optimization problem as the OASPL across various AOAs, with aerodynamic and geometric constraints. We introduce a variational autoencoder to enhance the shape parametrization of airfoils, to reduce the optimized parameters, and to generate a large family of airfoils. 

The variational autoencoder finds a large family of optimal geometries, whereas the classic CST parametrization converges to very similar shapes. Both methods find airfoils with comparable objective functions, but the variational autoencoder is superior in better satisfying the constraints with improved aerodynamic properties. Compared to the baseline geometry (the NACA0012 airfoil), the autoencoder-based optimized airfoil reduces by $3\%$ the overall sound pressure level (about 1.7dBA in Overall A-weighted Sound Pressure Level, OASPL), while improving aerodynamic performance. We conclude that the aeroacoustic optimization framework is a very useful tool for wind turbine airfoil design.

\section{Acknowledgements}
Laura Botero-Bolívar, Leandro de Santana, Oscar Mariño, Eusebio Valero thank the European Union Horizon 2020 Research and Innovation Program under the Marie Sklodowska-Curie grant agreement No 860101 for the zEPHYR project.
Esteban Ferrer would like to thank the support of the Spanish Ministry MCIN/AEI/10.13039/501100011033 and the European Union NextGenerationEU/PRTR for the grant ``Europa Investigación 2020'' EIN2020-112255, and also the Comunidad de Madrid through the call Research Grants for Young Investigators from the Universidad Politécnica de Madrid. 
Finally, all authors gratefully acknowledge the Universidad Politécnica de Madrid (www.upm.es) for providing computing resources on Magerit Supercomputer.

\appendix
\section{Parameters in Amiet's theory}\label{sec:appendix_amiet_param}

\subsection{Spanwise correlation length}\label{Sec: lambda}
A common methodology to obtain the spanwise correlation length is using Corcos' model~\cite{corcos1964}, which depends on an empirical constant. This constant can be obtained by fitting the coherence between two sensors located along the span of the airfoil; however, it is usually assumed as a constant value. The spanwise correlation length is related to the distance in the spanwise direction where there is a high level of coherence. It can also be related to the size of the turbulent structures in the spanwise direction. Theoretically, the spanwise correlation length is defined as the integral in the entire span of the coherence between two points in the spanwise direction. It can be calculated from the Corcos' model for the coherence between two spanwise points as:
\begin{equation}\label{Eq:corlen}
 \Lambda_{z\mid PP} = b_c\frac{\overline{U_c}}{\omega}   
\end{equation}
where $b_c$ is the Corco's constant equal to 1.4.

\subsection{Wall-pressure spectrum}\label{sec: WPS}
The wall pressure fluctuations in the vicinity of the trailing edge are the near-field noise source of the far-field noise emitted by an airfoil with an attached and turbulent boundary layer. Therefore, it might be the most important input of Amiet's model and critical for geometry optimization procedures. Therefore, a detailed discussion about the obtaining of this quantity is explained in Section~\ref{sec: WPS}. They are caused by velocity fluctuations related to the small structures near the airfoil wall and the larger structures in the outer part of the boundary layer~\cite{Devenport}. The smaller structures scale with viscous length ($u_\tau/\nu$) and are responsible for the high frequency range of the wall pressure spectrum, whereas the larger turbulent structures scale with the boundary layer thickness ($\delta$) and are responsible for the low frequency range of the wall pressure spectrum. The mid-frequency range scales with both types of structures. Based on this theory, several semiempirical models have been proposed to model the wall pressure spectrum, based on experimental measurements of flat plates or airfoils with pressure gradient~\cite{goody2004, chase1980, rozenberg2012, kamruzzaman2015semi, hu2016, lee2005modeling}. Other model is the one proposed by Blake~\cite{blake2017mechanics} which is based on physical analysis by solving the Poisson's equation, which governs the relationship between the velocity fluctuations across the boundary layer and the wall-pressure fluctuations in the airfoil wall. The model proposed by Blake~\cite{blake2017mechanics} was later extended by Parchen~\cite{parchen1998progress} and is now known as the TNO-Blake model. The TNO-Blake model incorporates more physics and takes into account several turbulence quantities, and not only integral quantities as in semiempirical models. For this reason, we retain this model to perform the optimization procedures presented in this work. 

The TNO-Blake model is proposed based on the solution of the Poison equation that results from the divergence of the Navier-Stokes equation simplified by the continuity equation. The model can be applied to calculate the wall-pressure spectrum close to the trailing edge, however, the turbulence quantities are calculated assuming that the discontinuity of the trailing edge does not affect the turbulence. Furthermore, the model assumes that the turbulent pressure and velocity field are spatially homogeneous and stationary on time. The wall pressure spectrum can be calculated in the wavenumber-frequency domain by applying the Fourier transform in time-space Fourier transform. Therefore, the wall pressure spectrum would depend on the wavenumber in the chordwise direction ($\kappa_x$) and spanwise direction ($\kappa_z$). Integration in $\kappa_z$ can be conducted by introducing the spanwise correlation length that depends on the frequency ($\Lambda_{z|pp3}(\omega)$.  In this research, we follow the TNO-Blake model extension proposed by Stalnov et al. ~\cite{stalnov2016towards}, which also considers the anistropy of the turbulence by incorporating the stretching parameters. 

The single-point wall pressure spectrum can be calculated as follows:
\begin{equation}\label{Eq:TNO}
\begin{small}
\Pi_\mathrm{pp}(\omega) = \frac{4\pi \rho^2}{\Lambda_{z\mid PP}(\omega)}\int_{0}^{\delta} \Lambda_{y\mid vv}(y) U_c(y)\left[ \frac{\partial U(y)}{\partial y}\right]^2 \frac{\bar{u_y^2}(y)}{U_c^2(y)}\phi_{vv}(\kappa_x,\kappa_z=0)e^{-2\mid \kappa \mid y}dy
\end{small}
\end{equation}
where $\delta $ is the thickness of the boundary layer, $\Lambda_{y\mid vv}(y)$ is the integral length scale across the boundary layer in the $y$ direction, i.e., in the direction of the boundary layer, perpendicular to the airfoil wall. $U_c(y)$ is the convection velocity across the boundary layer, which is assumed to be between 0.6 and 0.8 the velocity and the stream-wise velocity. $U(y)$ is the velocity at each position along the boundary layer, $u_y$ is the velocity fluctuation in the $y$ direction. $\phi_{vv}$ is the turbulence spectrum of the velocity fluctuations in the $y$-direction. $\mid\kappa\mid$ is the total wavenumber. Here, only the wave number in the chordwise direction is considered, thus, $\mid\kappa\mid = \kappa_x(y) = \omega/U_c(y)$. \ref{sec:appendix_BL_parameters} addresses the modeling of the boundary layer mean flow characteristics and turbulence quantities across the boundary layer, which are calculated using XFOIL, as discussed in the next section. 

\subsection{Boundary layer parameters}\label{sec:appendix_BL_parameters}
The following sections detail the relations between the aerodynamic parameters and the aeroacoustic model. 

\subsubsection{Boundary layer thickness}
The boundary layer thickness can be calculated with the empirical relation proposed by Drela~\cite{drela1989xfoil}:
\begin{equation}\label{eq: delta}
\delta = \theta \left (3.15 + \frac{1.75}{\frac{\delta^*}{\theta}+1} \right )+\delta^*
\end{equation}
where $\delta^*$ is the boundary layer displacement thickness and $\theta$ is the momentum thickness, which are obtained from XFOIL simulations. 

\subsubsection{Mean velocity profile}
The mean velocity across the boundary layer is modeled by the universal von-Kármán log of the wall coupled with the Coles' wake factor ($\Pi_w$) that accounts for the pressure gradient~\cite{coles1956law}, as:
\begin{equation}\label{eq: mean_velocity_profile}
\begin{split}
\frac{U(y)}{u_\tau}& = \frac{1}{\kappa}\log \left( y^+ \right) + B + \frac{2\Pi_w}{\kappa}\sin^2\left( \frac{\pi y}{2\delta} \right) \; \; \; \; \; y^+ \leq 5\\
\frac{U(y)}{u_\tau}& = y^+ \; \; \; \; \; \; \; \; \; \; \; \; \; \; \;\; \; \; \; \; \; \; \; \; \;\; \; \; \; \;\; \; \; \; \;\; \; \; \; \;\; \; \; \; \;y^+ < 5\\
\Pi_w &= 0.8 \left( \frac{\delta^* }{\tau_w}\frac{\mathrm{d} C_p}{\mathrm{d} x} + 0.5 \right)^{3/4} \\
u_\tau &= \sqrt{\frac{1}{2}c_fU_\infty ^2}
 \end{split}   
\end{equation}
where $u_\tau$ is the friction velocity, k is the von Kárman constant (= 0.38), B is the level constant (= 5), $y^+$ is the non-dimensional distance from the wall, $y^+ = yu_\tau/\nu$, $C_p$ the pressure coefficient, $\tau_w$ is the local wall shear stress equal to $\tau_w = 0.5\rho U_\infty ^2c_f$, and $c_f$ is the local friction coefficient. For the calculation of the mean velocity across the boundary layer, $\frac{\mathrm{d} C_p}{\mathrm{d} x}$ and $c_f$ are obtained from XFOIL.

\subsubsection{Turbulence intensities}
To calculate the velocity fluctuations in the streamwise direction, the model proposed by Alfredson et al.~\cite{alfredsson2011new} is used, as shown in Eq.~\ref{Eq:vel_fluc}, where $\gamma = 64$, $a = 0.2909$, and $b = 0.2598$. These values were obtained empirically for a zero pressure gradient boundary layer. To calculate the velocity fluctuations in the other direction, the strecthing parameters for anisotropic turbulence are adopted, i.e. $\beta_x = 1$, $\beta_y = 1/2$, $\beta_z = 3/4$. Thus, $\overline{u_y^2} = \beta_y \overline{u_x^2}$ and $\overline{u_z^2} = \beta_z \overline{u_x^2}$.
\begin{equation}\label{Eq:vel_fluc}
\begin{split}
    \frac{u_x(y)}{U(y)} &= \left( a  - b \left(\frac{U(y)}{U_\infty} \right ) \right)Q\\
    Q &= 1-e^{-\gamma\left (1-\frac{U(y)}{U_\infty}\right )}
\end{split}
\end{equation}

\subsubsection{Integral length scale}
The length scale relevant for wall pressure fluctuations is the one in the transversal direction of the boundary layer, i.e. the vertical size of the eddies ($\Lambda_{y\mid vv}(y)$). It can be calculated using the model of Schilichting~\cite{schlichting1961} using the concept of mixing length ($l_{\mathrm{mix}}$):
\begin{equation}\label{Eq:lambda}
\begin{split}
   \Lambda_{y\mid vv}(y) &= \frac{l_{\mathrm{mix}}}{k} \\
   l_{\mathrm{mix}} & =\frac{0.085\delta \tanh{\left ( \frac{k}{0.085}\left ( \frac{y}{\delta} \right) \right )}}{\sqrt{1+B \left (y/\delta \right )^6}}
\end{split}
\end{equation}

The integral length scale in the streamwise direction can be calculated assuming istropic turbulence, i.e., $\Lambda_{x\mid uu}(y) = 2\Lambda_{y\mid vv}(y)$

\subsubsection{Velocity spectrum}
The calculation of the velocity spectrum is based on the von Kármán model for the energy spectrum of isotropic turbulence integrated over $\kappa_y$. The incorporation of the stretching parameters accounts for the anisotropy, as shown in Eq.~\ref{Eq:phi_vv}, where $\kappa_e$ is the wavenumber of the eddies that contains most of the energy, therefore related to the streamwise integral length scale.
\begin{equation}\label{Eq:phi_vv}
\begin{split}
 \phi_{vv}& = \frac{4}{9\pi}\frac{\beta_x\beta_z}{\kappa_e^2}\frac{\left (\beta_x\kappa_x/\kappa_e \right )^2 + \left ( \beta_z \kappa_z/\kappa_e\right )^2}{\left [ 1 + \left (\beta_x\kappa_x/\kappa_e \right )^2 + \left ( \beta_z \kappa_z/\kappa_e\right )^2 \right ]^{7/3}}   \\
 \kappa_e(y) & = \frac{\sqrt{\pi}}{\Lambda_{x\mid uu}}\frac{\Gamma(5/6)}{\Gamma(1/3)}
 \end{split}
\end{equation}

 \bibliographystyle{elsarticle-num} 
 \bibliography{cas-refs}





\end{document}